\documentclass[fleqn,usenatbib]{mnras}

\usepackage[T1]{fontenc}
\usepackage{pdflscape}

\DeclareRobustCommand{\VAN}[3]{#2}
\let\VANthebibliography\thebibliography
\def\thebibliography{\DeclareRobustCommand{\VAN}[3]{##3}\VANthebibliography}

\usepackage{graphicx}	% Including figure files
\usepackage{amsmath}	% Advanced maths commands
\usepackage{amssymb}	% Extra maths symbols

\usepackage{gensymb}

\def \aj {AJ}
\def \mnras {MNRAS}
\def \apj {ApJ}

\def \apjl {ApJL}
\def \aap {A\&A}
\def \nat {Nature}
\def \araa {ARAA}
\def \pasp {PASP}

\usepackage{newtxtext,newtxmath}
\title[The radio jet of the Rapid Burster]{The variable radio jet of the accreting neutron star the Rapid Burster}

\author[J. van den Eijnden, D. Robins, R. Sharma et al.]{J. van den Eijnden$^{1}$\thanks{jakob.van-den-eijnden@warwick.ac.uk}\thanks{These authors contributed equally to this work}, D. Robins$^{1}$\footnotemark[2], R. Sharma$^{1}$\footnotemark[2], C. S\'anchez-Fern\'andez$^{2}$, T. D. Russell$^{3}$, \and N. Degenaar$^{4}$, J. C. A. Miller-Jones$^{5}$, and T. Maccarone$^{6}$
\\
$^{1}$Department of Physics, University of Warwick, Coventry CV4 7AL, UK\\
$^{2}$ Science Operations Department, European Space Astronomy Centre (ESA/ESAC), Madrid, Spain \\
$^{3}$ INAF, Istituto di Astrofisica Spaziale e Fisica Cosmica, Via U. La Malfa 153, I-90146 Palermo, Italy \\
$^{4}$ Anton Pannekoek Institute for Astronomy, University of Amsterdam, Amsterdam, The Netherlands \\
$^{5}$ International Centre for Radio Astronomy Research, Curtin University, Perth, Western Australia, Australia \\
$^{6}$ Department of Physics and Astronomy, Texas Tech University, Lubbock, TX, USA
}

\date{Accepted XXX. Received YYY; in original form ZZZ}

\pubyear{2023}

\begin{document}
\label{firstpage}
\pagerange{\pageref{firstpage}--\pageref{lastpage}}
\maketitle

% Abstract of the paper
\begin{abstract}
The Rapid Burster is a unique neutron star low-mass X-ray binary system, showing both thermonuclear Type-I and accretion-driven Type-II X-ray bursts. Recent studies have demonstrated how coordinated observations of X-ray and radio variability can constrain jet properties of accreting neutron stars -- particularly when the X-ray variability is dominated by discrete changes. We present a simultaneous VLA, \textit{Swift}, and \textit{INTEGRAL} observing campaign of the Rapid Burster to investigate whether its jet responds to Type-II bursts. We observe the radio counterpart of the X-ray binary at its faintest-detected radio luminosity, while the X-ray observations reveal prolific, fast X-ray bursting. A time-resolved analysis reveals that the radio counterpart varies significantly between observing scans, displaying a fractional variability of $38 \pm 5$\%. The radio faintness of the system prevents the robust identification of a causal relation between individual Type-II bursts and the evolution of the radio jet. However, based on a comparison of its low radio luminosity with archival Rapid Burster observations and other accreting neutron stars, and on a qualitative assessment of the X-ray and radio light curves, we explore the presence of a tentative connection between bursts and jet: i.e., the Type-II bursts may weaken or strengthen the jet. The former of those two scenarios would fit with magneto-rotational jet models; we discuss three lines of future research to establish this potential relation between Type-II bursts and jets more confidently.
\end{abstract}


\section{Introduction}

Jets are a fundamental component of a wide range of accretion-driven and cataclysmic transients in the universe, from, e.g., forming stars \citep[see e.g.,][for reviews]{stellarjetreview1,stellarjetreview2} and active galactic nuclei \citep{blandford2019} to gamma-ray bursts \citep{kumar2015} and compact object mergers \citep{combi2023}. Different jet-launching objects may provide complementary constraints on the physics of jet formation, collimation, and impact, by measuring the dynamic, spectral, and geometric properties of the jet across various time and size scales. Within the Milky Way, relativistic jets may be studied in depth in X-ray binaries: systems wherein a compact object accretes from a non-degenerate or white dwarf donor star. Low-mass X-ray binaries (LMXBs), where the donor star is typically defined to be $<1 M_{\odot}$, have historically been the most common target for such studies: their relative proximity and time scales of variability -- from sub-second variability in their accretion flow to weeks to months-lasting outbursts -- provide the opportunity to study the evolution of their jets across accretion rate and accretion flow state. 

The population of known Galactic LMXBs is divided between systems hosting a black hole or a neutron star, as the compact object \citep{degenaarbahramian,avakyan2023,fortin2024}, with the latter forming the majority. Both types of LMXBs are known to launch jets, dominant at the low-frequency end of the spectral energy distributions (radio and mm bands), albeit with different typical luminosities \citep{fender2000,migliari06,vandeneijnden2021}. Their ‘compact jets’ show a coupling to the accretion flow, observed as a correlation between accretion-driven X-ray and the jet-driven radio luminosities \citep{hannikainen98,corbel03,gallo03,migliari06,tudor17}. In black hole systems, this correlation is found to be $\sim 20$ times radio brighter than in neutron star LMXBs at the same X-ray luminosity\citep{fender2001,gallo18}. Another essential difference between both types of LMXBs is the presence of a solid surface and potential rotating stellar magnetic field in neutron stars: both this surface and magnetosphere interact with and affect the accretion flow, especially in the inner regions where the radio jet is thought to originate \citep{blandford1982,parfrey2016}. 

Coordinated radio and X-ray monitoring throughout LMXB outbursts has revealed a wide range of phenomenological properties of the coupling between accretion flow and jet. These properties include the aforementioned X-ray -- radio luminosity correlation, radio flaring, and the jet quenching observed in soft-state black hole LMXBs \citep{vadawale2003,fender2004outburststates,russell2020}. Additionally, jet ejecta launched during periods of radio flaring have been revealed systematically \citep{mirabel1994}, particularly in recent years \citep{russel2019,carotenuto2021,wood2021}. Such studies are typically performed by conducting repeated observations at a cadence of days to weeks, which allows for tracking the average radio and X-ray behaviour between the different observing epochs. In recent years, however, these studies have been complemented by low-frequency variability studies at higher time resolution \citep[e.g.,][]{atetarenko2021,vandeneijnden2020}, opening a complementary avenue to study the geometry and dynamics of LMXB jets and moving beyond the small number of radio-bright sources where such studies were performed earlier \citep[e.g.,][]{pooley1997}. 

In particular, coordinated studies tracking both the accretion flow and jet at high time resolution have yielded new insights into the propagation of accretion variability into the outflow. Observations of a set of accreting stellar-mass black hole systems, have revealed variability in compact jets as they respond to mass-accretion-rate fluctuations observed in X-rays \citep[Cyg X-1 and MAXI J1820+070;][]{atetarenko2019_cygx1,atetarenko2021} and in the transient jets launched by V404 Cyg \citep{atetarenko2017,atetarenko2021}. In the neutron star LMXB Swift J1858.6-0814, \citet{vincentelli2023} presented the detection of a jet response, measured in dual-band radio observations, to instabilities in the accretion flow, as probed by fast IR observations. In particular, assuming the launch of discrete ejecta to coincide with each transition between different stages of the accretion flow instability, \citet{vincentelli2023} were able to reproduce the observed radio variability with a superposition of four launch epochs of outflowing ejecta. Their same approach was also shown to reproduce the jet variability observed in the black hole LMXB GRS 1915+105, which undergoes the same accretion flow instabilities \citep[][]{pooley1997}. 

The key to these multi-band variability studies lies in the combination of spectral and timing information, tracking outflowing material as it moves down the jet and shifts its emission to lower frequencies \citep{blandford1979}. The high-quality, multi-wavelength data sets required for these studies have restricted them predominantly to black hole systems. Apart from Swift J1858.6-0814, no other neutron star LMXB has shown accretion instabilities that could be associated with jet ejecta \citep{vincentelli2023}. The relative faintness of the compact jets from neutron star LMXBs has furthermore prevented detections of compact jet variability resulting from continuous accretion rate fluctuations. 

However, the presence of a stellar surface and rotating magnetic field in neutron star LMXBs offers a complementary approach to track the short-time-scale jet responses to the dynamics of the inner accretion flow. Firstly, accreted material on the neutron star surface may ignite in a runaway thermonuclear explosion, lasting several seconds followed by a cooling tail lasting tens of seconds, known as a Type-I burst \citep[e.g.,][]{galloway2021}. Type-I bursts generate soft X-ray photons that cool down the accretion flow via successive inverse-Compton scatterings \citep{kajava2017,sanchez2020} and cause brief increases of the accretion rate \citep{degenaar2018,fragile2018,fragile2020}. The brightness and short duration of Type-I bursts make them powerful signposts of the accretion flow changes they cause. Through an extensive and simultaneous X-ray and radio monitoring campaign, \citet{russell2024} recently found that the radio jets in two neutron star LMXBs (4U 1728-34 and 4U 1636-536) show bright radio flares in the minutes succeeding Type-I bursts. This jet response, delayed increasingly towards lower radio frequencies, yielded the first measurement of the compact jet speed in a neutron star LMXB: $v=0.38^{+0.11}_{-0.08}c$, which is consistent with the expected escape speed of a neutron star. 

The rotating neutron star magnetic field may also affect the inner accretion flow. In LMXBs, where the neutron stars typically are weakly magnetized ($B\lesssim10^9$ G), these interactions most often take the form of magnetic truncation of the inner flow \citep{ghosh1978,ludlam2024}. Gas is subsequently channelled onto the neutron star’s magnetic poles, leading to X-ray pulsations \citep[see e.g.,][for a recent review]{disalvo2023}. In rare cases, however, neutron star LMXBs may display Type-II X-ray bursts. While the nature of these bursts, which typically last between seconds and a minute \citep{bagnoli2015}, remains not fully understood, one common explanation lies in the disk-magnetosphere interaction: in the trapped disk model \citep{spruit1993,dangelo2010}, the accretion flow encounters a centrifugal magnetic barrier, when its inward pressure is balanced by the magnetic field slightly outside the radius where the disk’s rotation equals the neutron star spin (the ‘co-rotation’ radius). As material is trapped and accumulates, its inward pressure increases, pushing the magnetic barrier inwards until it crosses the co-rotation radius and overcomes the centrifugal barrier. The resulting increase in accretion rate manifests as a Type-II bursts. This trapped disk model naturally explains the observed relation between Type-II burst fluence and wait time between bursts \citep{bagnoli2015}; it is further consistent with the measurement of large disk truncations in the only two sources showing Type-II bursts : the Rapid Burster \citep[MXB 1730-335;][]{vandeneijnden2017} and the Bursting Pulsar \citep[GRO J1744-28;][]{degenaar2014}. In addition, it is consistent with the brief dips in X-ray emission seen before and after some Type-II, which could signal the depletion of the accretion flow \citep{bagnoli2015,younes2015,court2018}. On the other hand, a range of open questions remains regarding the nature of Type-II bursts; for instance, both the trapped disk and alternative models have difficulty to explain the rarity of the Type-II bursts amongst the neutron star LMXB population \citep[see e.g.][for a recent discussion]{lyutikov2023}. We therefore refer the reader to \citet{bagnoli2015} for a more extensive introduction into Type-II bursts, including alternative models. 

During Type-II bursts, observational evidence suggests that the accretion rate increases significantly \citep{bagnoli2015,court2018}, consistent with the trapped disk model. This change in accretion rate may, subsequently, lead to observable changes in the jets launched by the LMXB. The nature of this response depends on the mechanism underlying the neutron star jet launch itself: in the classic \citet{blandford1982} model, the rotating inner accretion flow is responsible for tangling up the toroidal disk magnetic field, leading to the launch of jets. In more recent models, introduced by \citet{parfrey2016} and explored numerically in \citet{das2022}, \citet{parfrey2023}, \citet{berthier2024}, and \citet{das2024}, the spin and magnetic field of the neutron star instead provide the jet power, as the accretion flow opens up field lines to launch an outflow. In the former case, the significant truncation of the inner disk in between bursts may lead to a weaker jet, that brightens during bursts as more material reaches the jet-launch regions. In the latter type of model, the swift accretion of material trapped by the centrifugal barrier in between bursts may reduce the gas influx and weaken the jet; alternatively, it may not affect the jet substantially, if the power and matter are provided over a range of radii beyond the barrier. The detection of a potential jet response to Type-II bursts can therefore constrain both properties of the jet, such as its speed, and inform the underlying launch mechanism. 

Both the Rapid Burster and Bursting Pulsar are transient LMXBs. The radio counterpart of the Bursting Pulsar is not known, and due to its location close to the Galactic Centre and relatively long recurrence time between outbursts (five outbursts since 1995), limits on its radio emission are shallower than for other neutron star LMXBs \citep{russell2017,vandeneijnden2021}. The Rapid Burster, on the other hand, is a prolific transient LMXB, showing quasi-regular outbursts lasting weeks and recurring every $\sim$ 80 -- 105 days \citep{guerriero1999}. In the past $\sim 5$ decades, radio studies of the Rapid Burster or its environment have been carried out repeatedly at increasing sensitivity \citep{johnson1978,rao1980,grindlay1986,johnston1991,tudor2022}. Its radio counterpart was first confidently identified by \citet{moore2000}. They further assessed whether the radio counterpart responds to the Type-II bursts, but their findings were limited by the signal-to-noise ratio in establishing the presence of such a response. The Rapid Burster is located in the globular cluster Liller 1, at a distance of 7.9 kpc \citep{valenti2010}. Due to this crowded stellar environment and high absorption, no optical or infrared counterpart of the LMXB is known \citep{homer2001}. 

Making use of the quasi-predictable outburst occurance and evolution, we designed a coordinated radio and X-ray campaign of the Rapid Burster to investigate the effect of Type-II bursts on its radio jet at unprecedented sensitivity. In this coordinated campaign on 19 March 2020, the Karl G. Jansky Very Large Array (VLA), the \textit{Neil Gehrels Swift Observatory} \citep[\textit{Swift};][]{gehrels2004}, and the \textit{INTErnational Gamma-Ray Astrophysics Laboratory} \citep[\textit{INTEGRAL};][]{winkler2003} observed the LMXB simultaneously. In this paper, we present the radio and X-ray results of this campaign and search for a jet response to Type-II bursts, as a probe of both the neutron star jet launching and origin of Type-II bursts. 

\section{Observations and data analysis}

\subsection{2020 Campaign: VLA, \textit{Swift}, and \textit{INTEGRAL}}
\label{sec:data_2020}

To study the time-averaged and variable radio jet of the Rapid Burster, we carried out a coordinated, simultaneous observing campaign with the VLA, \textit{Swift}, and \textit{INTEGRAL} observatories. The observations were taken on 19 March 2020, where multi-wavelength data were collected simultaneously between 12:11:48 and 14:14:17 UTC, with the exception of a number of gaps for radio calibration and X-ray Earth occultation (see below). This overlap between observatories was limited by the duration of the radio observation. 

Triggered by the potential and expected onset of an outburst in all-sky monitoring by the \textit{Monitor of All-Sky X-ray Image} \citep[\textit{MAXI};][]{matsuoka2009} aboard the International Space Station (ISS), \textit{Swift} monitored the Rapid Burster (target ID 31360) with six short ($\sim0.4 - 0.5$ ks) observations between 3 and 19 March 2020. These observations confirmed the outburst and motivated the scheduling of the coordinated epoch, optimized to catch the target in a phase with a large number of Type-II bursts. During this coordinated epoch, \textit{Swift} observed the Rapid Burster in two consecutive exposures separated by an Earth occultation: ObsIDs 00031360171 and 00031360172, each lasting 1.6 ks and starting at 12:06:34 and 13:41:36 UTC, respectively. The X-ray Telescope (XRT) was employed in WT mode. 

We employed the online \textit{Swift} XRT products pipeline\footnote{\url{https://www.swift.ac.uk/user_objects/}} \citep{evans2007} to extract a combined, time-averaged X-ray spectrum from ObsIDs 00031360171 and 00031360172. We also extracted a light curve at a 1-second time resolution between 1 and 10 keV, in order to determine the presence of Type-II (or Type-I) bursts and measure their start and end times. Type-II bursts are mostly achromatic below 10 keV, as the increase in mass accretion rate predominantly affects the spectral normalisation \citep{bagnoli2015}; therefore, we do not extract separate burst and non-burst spectra. When analysing the X-ray spectrum, as described in Section \ref{sec:xrayanalysis}, we used \textsc{xspec} v12.13.0c \citep{xspecref}. We assumed abundances and cross-sections from \citet{wilms2000} and \citet{vern1996}, respectively, when accounting for interstellar absorption. We quote all measured X-ray fluxes and spectral parameters in this paper at their $1\sigma$ uncertainties; all reported fluxes are corrected for absorption. 

We observed the Rapid Burster with the VLA between 12:00:00 and 14:15:37 UTC on 19 March 2020 under program 20A-172. The first scan of the Rapid Burster started at 12:11:48. In total, we obtained 11 scans of the target field, each lasting 519 seconds, with the final scan finishing at 14:14:17. As primary and secondary calibrators, we observed 3C286=J1331+305 and J1744-3116, respectively. In order to optimize the simultaneous coverage in radio and X-rays, we observed the primary calibrator after eight target scans, during \textit{Swift}'s Earth occultation, before returning to the target. The observation was carried out at C band in 3-bit mode, yielding 4 GHz of bandwidth between 4 and 8 GHz. The data were collected with a dump time of 5 seconds; the array was in its C configuration. We used the Common \textsc{Astronomy Software Application} v6.5.5.21 \citep[\textsc{CASA};][]{casa2022} to further analyse the data. Using a combination of automated routines and manual inspection, we removed RFI from the uncalibrated measurement set, before applying standard routines to perform the calibration\footnote{Detailed flagging and calibration logs are included in this paper's reproduction package; see the Data Availability Statement.}. 

To analyse the time-averaged observation, we imaged the target field using \textsc{tclean}, where we used Briggs weighting with a robust parameter of zero to optimize the balance between increased sensitivity and minimising the effects of side-lobes in the telescope beam pattern as well as diffuse emission within the field. Due to the faintness of the target, we did not split the observation into multiple frequency bands. As discussed in Section \ref{sec:results_averaged}, we removed the inner baselines, covering the inner $3.5\times10^3\lambda$, from the measurement set before both the time-averaged and time-resolved analysis: this uv-plane cut optimizes the removal of diffuse emission present in the full image, without introducing a significant loss in sensitivity. To measure flux densities in the image plane, we used the \textsc{casa} tool \textsc{imfit} to fit an elliptical Gaussian source profile with its FWHM and orientation fixed to the size of the synthesized beam ($6.07\times1.98$ arcsec, position angle $2.39\degree$ East of North, in the full time-averaged observation; for images of subsets of the data, as discussed below, the synthesized beam changes shape).  The RMS sensitivity of each analysed image was measured in a source-free region close to the target position; the time-averaged observation reaches a sensitivity of $5.5$ $\mu$Jy.

To study the time-resolved properties of the Rapid Burster, on a range of time scales (introduced in Section \ref{sec:results_resolved}), we measured the source's flux density in the uv-plane via the \textsc{CASA} tool \textsc{uvmodelfit}. For this purpose, we first identified all other sources in the time-averaged image, after removing the inner baselines. We measured their positions and flux densities using \textsc{imfit} before using the \textsc{ft} tool to define a model containing these background sources. Using \textsc{uvsub}, we created a measurement set where these sources are subtracted, after which we used \textsc{phaseshift} to shift the pointing centre of that measurement set at the position of the Rapid Burster. The resulting measurement set was used for the uv-plane fitting at the different employed time resolutions. In order to assess whether any observed variability is intrinsic to the target, we repeated all steps above to create a second measurement set, replacing the Rapid Burster by a check source of similar flux density (no substantially brighter sources are present in the field) located at $17\text{h } 33\text{m } 10.75\text{s}$, $-33^{\rm o} 24\text{' } 26\text{"}$. As discussed in Section \ref{sec:results_resolved}, we also checked the validity of these uv-plane results with image-plane comparisons.

\textit{INTEGRAL} observations of the Rapid Burster were executed between 10:22:03 and 18:04:29 UTC on 19 March 2020. The observations were performed using the hexagonal (HEX) dithering mode, which consists of one pointing on the source, and six pointings surrounding it, arranged an hexagonal shape. The angular distance between consecutive pointings is $2.17\degree$, and the pointing duration 1872 seconds. We used data from the Joint European X-ray Monitor (JEM-X) \citep{Lund03}. JEM-X consists of two identical units X1 and X2, is sensitive in the 3–35 keV range where it provides an angular resolution of 3$\arcmin$. The data were reduced  with the INTEGRAL Offline Science Analysis (OSA) v11.0 \citep{Courvoisier2003}, provided by the INTEGRAL Science Data Center (ISDC). The data were analysed using standard procedures. The light curves of the Rapid Burster were generated in in the 3–25 keV band, with a time resolution of 5 seconds.

\subsection{Archival radio and X-ray observations}

\begin{table*}
  \begin{centering}
  \begin{tabular}{lccccccc}
   \hline
Date & $\nu_{\rm R}$ & $F_{\rm R}$ & $L_{\rm R}$ [5 GHz] & $R_{\rm PCA}$ & $F_{\rm X}$ & $L_{\rm X}$ & Fluence in bursts \\
& [GHz] & [$\mu$Jy] & [erg/s] & [cts/s] & [erg/s/cm$^2$] & [erg/s] & [\%] \\
    \hline
\multicolumn{8}{c}{\textit{Archival data}}     \vspace{1mm}\\
1996 Nov 6 & 8.4 & $370\pm45$ & $(1.4\pm0.2)\times10^{29}$ & $2830 \pm 150$ & $(7.2\pm1.7)\times10^{-9}$ & $(5.3\pm1.2)\times10^{37}$
 & $0$ \\ 
1996 Nov 11 & 4.9 & $190\pm45$ & $(7.1\pm1.7)\times10^{28}$ & $1660 \pm 100$ & $(4.2\pm1.0)\times10^{-9}$ & $(3.1\pm0.7)\times10^{37}$ & $0$ \\ 
1997 Jun 29 & 4.9 & $210\pm45$ & $(7.8\pm2.6)\times10^{28}$ & $3680 \pm 200$ & $(9.3\pm2.2)\times10^{-9}$ & $(7.0\pm1.6)\times10^{37}$ & $0$ \\ 
1997 Jul 24 & 8.4 & $<90$ & $<3.4\times10^{28}$ & $210 \pm 100$& $(5.3\pm2.8)\times10^{-10}$ & $(4.0\pm2.1)\times10^{36}$ & $>78$ \\ 
1998 Feb 19 & 8.4 & $<90$ & $<3.4\times10^{28}$ & $360 \pm 120$ & $(9.1\pm3.7)\times10^{-10}$ & $(6.8\pm2.7)\times10^{36}$ & $100$ \\ 
2015 Jan 06 & 7.25 & $<13.2$ & $<4.9\times10^{27}$ & -- & -- & -- & -- \\
\hline
\multicolumn{8}{c}{\textit{2020 campaign}}\vspace{1mm}\\
2020 Mar 19 & 6.0 & $58.1 \pm 5.5$ & $(2.2 \pm 0.2)\times10^{28}$ & -- & $(8.0\pm2.5)\times10^{-9}$ & $(6.0 \pm 1.9)\times10^{37}$ & 85 \\ 
\hline
  \end{tabular}
  \caption{Overview of the observation-averaged X-ray and radio properties of the Rapid Burster in archival observations (based on \citet{moore2000} and \citet{tudor2022} and the 2020 campaign (this work). The columns list the epoch date, radio observing frequency $\nu_{\rm R}$, radio flux density $F_{\rm R}$, inferred radio luminosity $L_{\rm R}$  at 5 GHz, \textit{RXTE}/PCA count rate (summed over five PCUs) $R_{\rm PCA}$, unabsorbed X-ray flux $F_{\rm X}$, inferred unabsorbed X-ray luminosity $L_{\rm X}$, and fraction of fluence in the X-ray observation detected during Type-II bursts. The X-ray flux and luminosity are bolometric (archival data) or in the 0.5-10 keV band (2020 campaign); the former include a systematic uncertainty capturing the effect of converting PCA rates to fluxes. Upper limits are reported at $3\sigma$. All listed epochs are based on simultaneous X-ray and radio observations.}
  \label{tab:archival}
  \end{centering}
\end{table*}

\label{sec:data_archival}

Since its discovery, the Rapid Burster has been the subject of multiple coordinated X-ray and radio observing campaigns. The lower sensitivity of these observations compared to the 2020 campaign, particularly in the radio band, yielded relatively unconstraining limits on the response of the radio jet to the presence of X-ray bursts. However, the time-averaged properties of these observations, as well as information on the prominence of Type-II bursts during these campaigns, provide crucial context for the interpretation of the results from 2020 campaign. In our analysis, we do not include the radio observations\footnote{We also do not include the results from \citet{calla1980}, which claims the detection of radio bursts with flux densities of $400$-$600$ Jy, as such remarkably large flux densities have not been confirmed or observed since.} presented in \citet{johnson1978}, \citet{rao1980}, \citet{johnston1991}, and \citet{grindlay1986}: these studies do not confidently detect a radio counterpart of the Rapid Burster, nor a radio response to its Type-II bursts. More recent studies \citep[e.g.,][]{moore2000} demonstrate that, indeed, the radio sensitivity in these studies was insufficient to detect this radio counterpart. In addition, these studies do not present X-ray flux measurements, but at most discuss the number of Type-II bursts, which further limits a comparison with later studies. 

\citet{moore2000} presented the detection of the Rapid Burster counterpart using VLA, the Australia Telescope Compact Array (ATCA), and James Clerk Maxwell Telescope/Sub-millimeter Common User Bolometer Array observations. As we discussed in more detail later, the Rapid Burster did not show Type-II bursts during the radio-detected epochs. Their campaign includes five epochs with coordinated \textit{Rossi X-ray Timing Explorer} (\textit{RXTE}) observations, providing a count rate measurement from its Proportional Counter Array (PCA) instrument. To convert these reported X-ray count rate measurements to X-ray fluxes, we adopt the conversion of PCA count rate to bolometric X-ray flux derived by \citet{bagnoli2015}: 1 count/s/PCU equals $(1.26 \pm 0.07)\times10^{-11}$ erg/s/cm$^2$, where the error is statistical; in our further calculations, we also include a $23$\% systematic uncertainty in the conversion, as further derived by \citet{bagnoli2015}, by quadratically combining both uncertainties\footnote{Generally, PCA observations can consist of up to five Proportional Counter Units, or PCUs; in all five observations discussed here, five PCUs were active, as shown by \textit{RXTE}'s \textit{mission-long data products} for the Rapid Burster. The PCA count rates reported by \citet{moore2000} are the sum of all PCUs and have therefore been divided by five in our analysis.}. With this systematic uncertainty, we also aim to capture the difference between this bolometric X-ray flux and the 0.5-10 keV X-ray flux typically calculated for other X-ray binaries. In the radio band, three of the five coordinated observations in \citet{moore2000} are reported at a single frequency of $8.4$ GHz: one detection and two non-detections. For those three epochs, we assume a flat spectrum to convert the reported flux density (limit) to a $5$ GHz luminosity. In the other two radio epochs, with detections at both $4.9$ and $8.4$ GHz, we adopt the former, with its uncertainty, for this calculation. The measured values from \citet{moore2000} and derived X-ray fluxes and X-ray/radio luminosities are listed in Table \ref{tab:archival}. 

\begin{figure}
\includegraphics[width=\columnwidth]{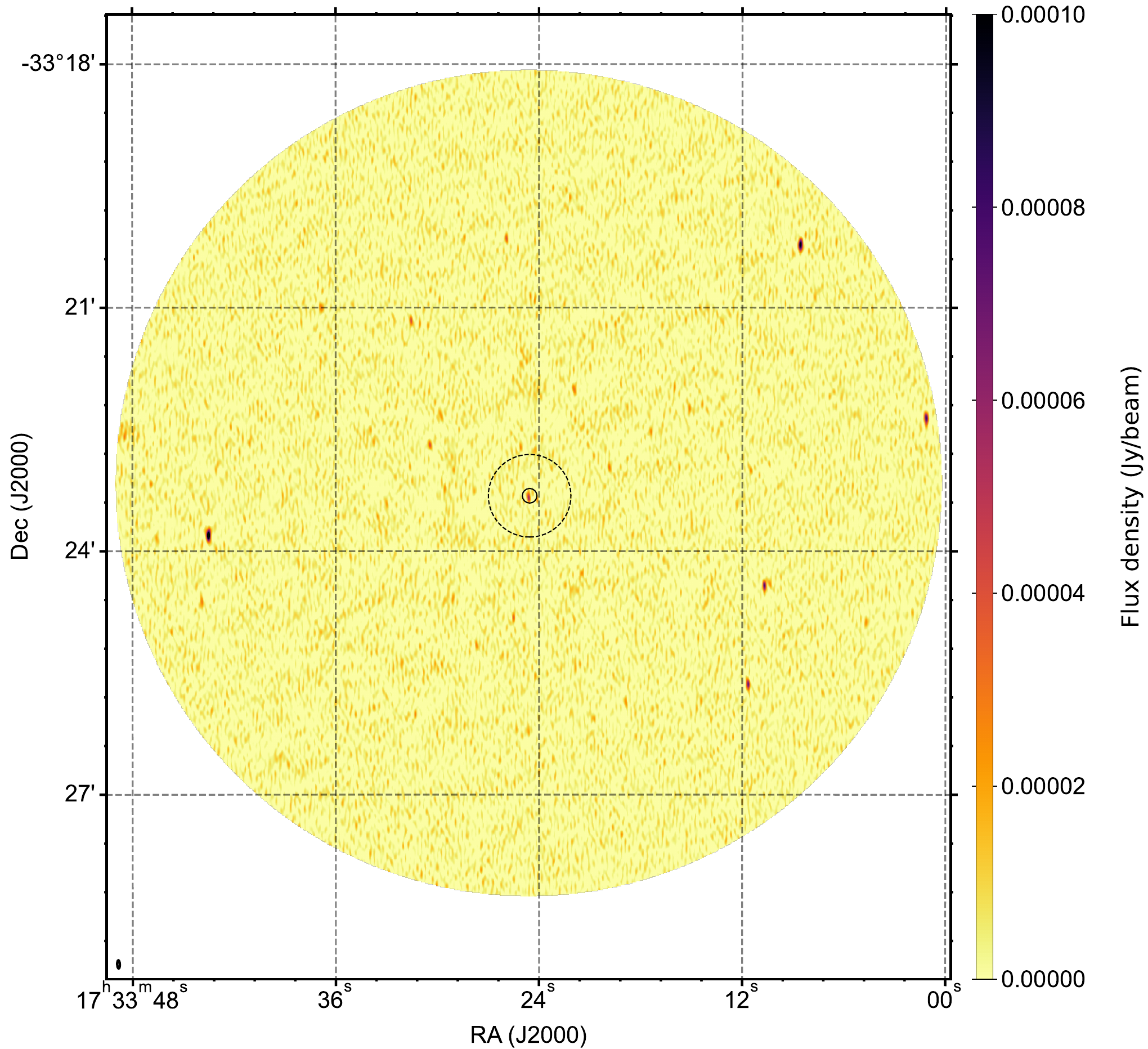}
 \caption{The observation-averaged, full-band, complete-FoV radio image of the Rapid Burster in Liller, generated from the 2020 VLA observation. The radio counterpart of the Rapid Burster is detected as the point source at the centre of the image. The synthesized beam is shown in the bottom left corner. The half-light and core radius of Liller 1 are indicated by the dashed and solid circle, respectively \citep{tudor2022}.}
 \label{fig:fullfield}
\end{figure}

More recently, \citet{tudor2022} presented deep ATCA observations of the globular cluster Liller 1, as part of the MAVERIC survey of such clusters. The observations, taken in 2015, show the absence of compact radio sources at the core of the cluster, i.e., consistent with the position of the Rapid Burster, down to an RMS sensitivity of $4.4$ $\mu$Jy/beam at $7.25$ GHz (combining the 5.5 and 9 GHz ATCA basebands). No pointed X-ray observations where taken simultaneous with or close to this ATCA epoch. However, inspection of the long-term X-ray monitoring light curve from \textit{MAXI} reveals no evidence for X-ray activity or an outburst. During quiescence, the \textit{MAXI} data of the Rapid Burster are dominated by contaminating flux from the persistent accreting neutron star 4U 1728-34, separated by an angular distance of $\sim 0.5\degree$. Therefore, the \textit{MAXI} light curve and spectrum do not provide constraining upper limits on the Rapid Burster's X-ray flux in quiescence. However, the lack of evidence of an outburst -- which are routinely detected by \textit{MAXI} above the persistent emission from 4U 1728-34 -- and the radio limit below our 2020 VLA detection can be used to associate the 2020 radio emission to the Rapid Burster (see Section \ref{sec:results_averaged}). 

\section{Results}

\subsection{Time-averaged analysis}
\label{sec:results_averaged}

% Radio paragraph: what average Lr do we measure? How does this compare to archival values? What does this mean?

\begin{figure*}
\includegraphics[width=\textwidth]{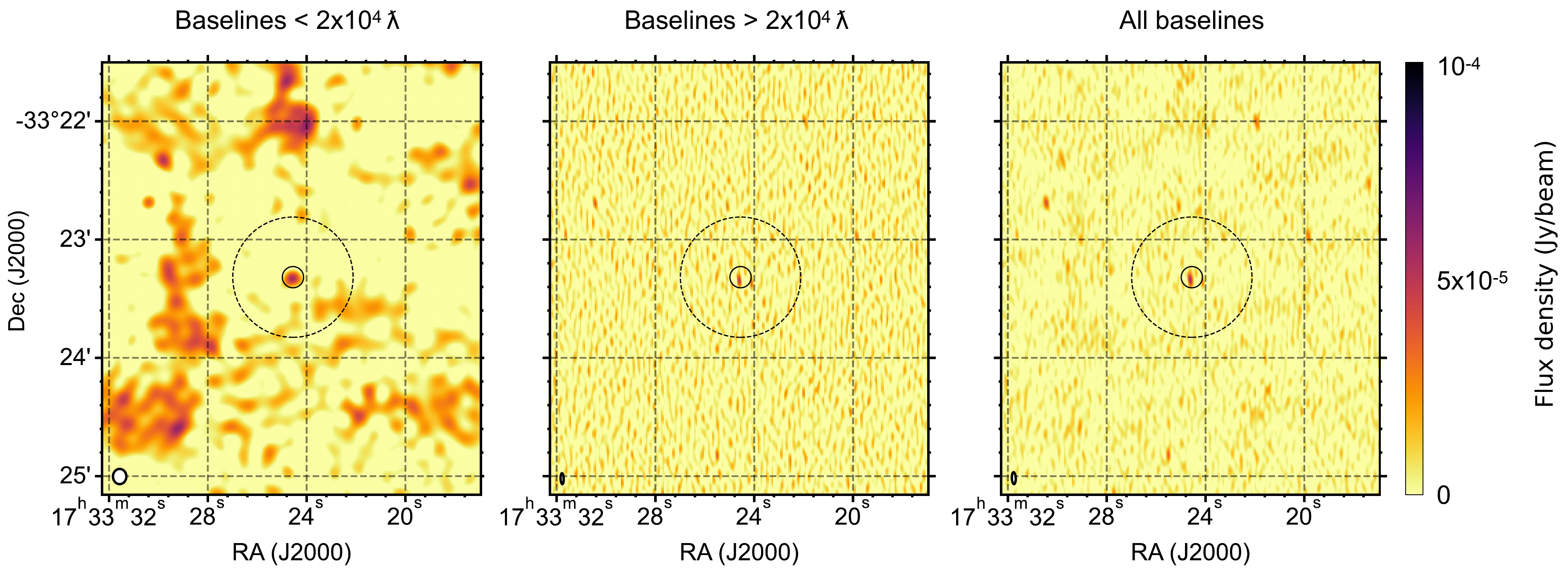}
 \caption{A zoom of the central $4\times4$ arcminutes around the position of the Rapid Burster in the time-averaged VLA observation. The three panels show the result of imaging the measurement set with different baseline selections: only baselines within $2\times10^4\lambda$ (left), only baselines outside of $2\times10^4\lambda$ (middle), and all baselines (right; e.g., the same as Figure \ref{fig:fullfield}). All three panels are shown on the same color scale. The lowest spatial resolution image on the left, as indicated by the size of the synthesized beam in the bottom left corner of each panel, shows the presence of extended emission on scales beyond the half-light radius of Liller 1; this half-light radius is indicated by the dashed circle, while the solid circle shows the core radius of the cluster \citep{tudor2022}.}

\label{fig:zooms}
\end{figure*}

In Figure \ref{fig:fullfield}, we show the entire field of view of the time-averaged 2020 VLA observation of the Rapid Burster and its surroundings. A faint point source can be identified in the centre of the image, which we will discuss in detail below. In addition to a number of other point sources across the field, the image also reveals apparent structure on scales larger than the image resolution: the residual structure of the background surrounding these point sources appears to deviate from a pure noise profile with the beam size as the characteristic size scale. To further demonstrate the presence of such extended emission in the field, we show three zooms of the field in Figure \ref{fig:zooms}. The left and middle panels show the image, re-generated by selecting only baselines smaller or larger than $2\times10^4\lambda$, respectively. The left panel, as expected with reduced spatial resolution, confirms the presence of extended emission on scales beyond the half-light radius of Liller 1 \citep[indicated by the dashed line;][]{tudor2022}. The middle, high-resolution panel, instead shows only the point sources seen in Figure \ref{fig:fullfield}. For completeness and comparison, the right panel shows the same zoom in the image without any baselines cuts applied. Due to the presence of diffuse emission visible on the shortest baselines, we apply a $3.5\times10^3\lambda$ baseline cut in all further analysis of the VLA observation: we converged to this value by trialing a range of uv-plane cuts, optimizing the trade-off between the detection of diffuse emission and loss of sensitivity. 

As shown in Figure \ref{fig:zooms}, a clear but faint radio point source is present at the centre of Liller 1. Using \textsc{imfit}, we measure this source's flux density as $58.1 \pm 5.5$ $\mu$Jy. We measure a source position of:

\begin{align*}
    &\text{RA (J2000)} = 17\text{h } 33\text{m } 24.61\text{s } \pm 0.01\text{s} \\
    &\text{Dec (J2000)} = -33^{\rm o} 23\text{' } 20.1\text{" } \pm 0.6\text{"}
\end{align*}

\noindent where the uncertainties reflect $10$\% of the beam size when imaging the field using all baselines. This position is consistent within $1\sigma$ uncertainties with the radio position of the Rapid Burster reported in \citet{moore2000}: the offset between the two positions is $0.3$ arcseconds in declination, and $0$ (at the quoted number of significant digits) in right ascension. In Table \ref{tab:coords}, we list the coordinates and 6-GHz flux densities of the other significantly-detected point sources in the image as well. In order to constrain the spectral index, we also image the full observation in two sub-bands, from $4-6$ and $6-8$ GHz. We measure flux densities of $63\pm8$ $\mu$Jy and $55\pm8$ $\mu$Jy, respectively, where the uncertainty again reflects the image RMS sensitivity. These values imply a poorly constrained spectral shape of $\alpha = -0.4 \pm 1.3$, consistent with the full range of expected indices for LMXB jets.

\begin{table}
\begin{centering}
\begin{tabular}{lll}
\hline
Right Ascension & Declination & 6 GHz flux density \\ \hline
17h 33m 24.61s $\pm$ 0.01s &$-$33$\degree$ 23' 20.1" $\pm$ 0.6" & $58.1 \pm 5.5$ $\mu$Jy \\
\hline
17h 33m 43.50s $\pm$ 0.01s &$-$33$\degree$ 23' 48.3" $\pm$ 0.6" & $155.5 \pm 5.5$ $\mu$Jy \\
17h 33m 30.44s $\pm$ 0.01s &$-$33$\degree$ 22' 41.4" $\pm$ 0.6" & $42.5 \pm 5.5$ $\mu$Jy \\
17h 33m 25.92s $\pm$ 0.01s &$-$33$\degree$ 20' 08.9" $\pm$ 0.6" & $40.1 \pm 5.5$ $\mu$Jy \\
17h 33m 25.51s $\pm$ 0.01s &$-$33$\degree$ 24' 49.0" $\pm$ 0.6" & $31.5 \pm 5.5$ $\mu$Jy \\
17h 33m 24.11s $\pm$ 0.01s &$-$33$\degree$ 23' 21.4" $\pm$ 0.6" & $18.7 \pm 5.5$ $\mu$Jy \\
17h 33m 11.66s $\pm$ 0.01s &$-$33$\degree$ 25' 38.4" $\pm$ 0.6" & $77.6 \pm 5.5$ $\mu$Jy \\
17h 33m 10.70s $\pm$ 0.01s &$-$33$\degree$ 24' 25.5" $\pm$ 0.6" & $80.4 \pm 5.5$ $\mu$Jy \\
17h 33m 08.59s $\pm$ 0.01s &$-$33$\degree$ 20' 13.6" $\pm$ 0.6" & $127.3 \pm 5.5$ $\mu$Jy \\
17h 33m 01.16s $\pm$ 0.01s &$-$33$\degree$ 22' 21.7" $\pm$ 0.6" & $96.1 \pm 5.5$ $\mu$Jy \\
\hline
\end{tabular}
\caption{The significantly detected point sources in the 20A-172 VLA observation of the Rapid Burster in Liller 1. The Rapid Burster is listed as the first source. The flux densities are reported at 6 GHz. The positional uncertainties reflect $10\%$ of the beam.}
\label{tab:coords}
\end{centering}
\end{table}

However, despite this positional agreement, the 2020 VLA observation alone does not necessarily imply that the radio source corresponds to the Rapid Burster: the centre of Liller 1 presents a dense environment, where lower frequency and lower resolution observations have revealed emission that likely originates from an unresolved population of steep-spectrum radio pulsars \citep{fruchter1995}. The source revealed by the VLA in 2020 is fainter than all earlier radio detections of the Rapid Burster, and lies below the upper limits of the earlier radio non-detections discussed in \citet{moore2000}. Therefore, we need to consider the scenario that this faint source corresponds to a persistent, unresolved population of radio sources in the cluster. As discussed in Section \ref{sec:data_archival}, a deep ATCA observation, taken in between outbursts in 2015, did not detect any radio emission from the core of Liller 1 down to a $3\sigma$ limit of $13.2$ $\mu$Jy. This limit, significantly below the 2020 VLA detection and at similar frequency, rules out this scenario of an unresolved source population; we therefore conclude that this VLA source is the counterpart of the Rapid Burster, observed at its faintest detected radio luminosity.  

\label{sec:xrayanalysis} 
To measure the X-ray flux during the 2020 VLA observation, we fit the time-averaged \textit{Swift}/XRT spectrum observed simultaneously with the radio observation. We find that out of single component models, such as an absorbed thermal or power law spectrum, an absorbed accretion disc model (\textsc{tbabs*diskbb}) provides the best fit with $\chi^2_\nu = 1040/886 = 1.18$. Models with two simple components, however, provide statistically superior fits, with the best fit provided by a \textsc{tbabs*(bbody+po)} model at $\chi^2_\nu = 957/884 = 1.08$. As we aim to use the \textit{Swift}/XRT spectrum only to obtain a measurement of the X-ray flux, we refer the reader to earlier works for a more in-depth discussion on X-ray spectral modelling in the Rapid Burster, making use of spectra with longer exposures and broader energy coverage \citep[see e.g.][and references therein]{vandeneijnden2017}. For this best-fit model, we find an absorbing column density of $N_H = (6.8\pm0.4)\times10^{22}$ cm$^{-2}$, black body temperature and normalizations of $T_{\rm BB} = 1.84\pm0.04$ keV and $N_{\rm BB} = (2.7\pm0.1)\times10^{-2}$, and a power law index and normalization of $\Gamma = 3.7 \pm 0.3$ and $N_{\rm po} = 2.1^{+0.8}_{-0.6}$. Using the convolution model \textsc{cflux}, we measure an unabsorbed 0.5-10 keV X-ray flux of $(8.0\pm2.5)\times10^{-9}$ erg/s/cm$^2$, corresponding to a luminosity of $(6.0 \pm 1.9)\times10^{37}$ erg/s. 

\begin{figure*}
\includegraphics[width=\textwidth]{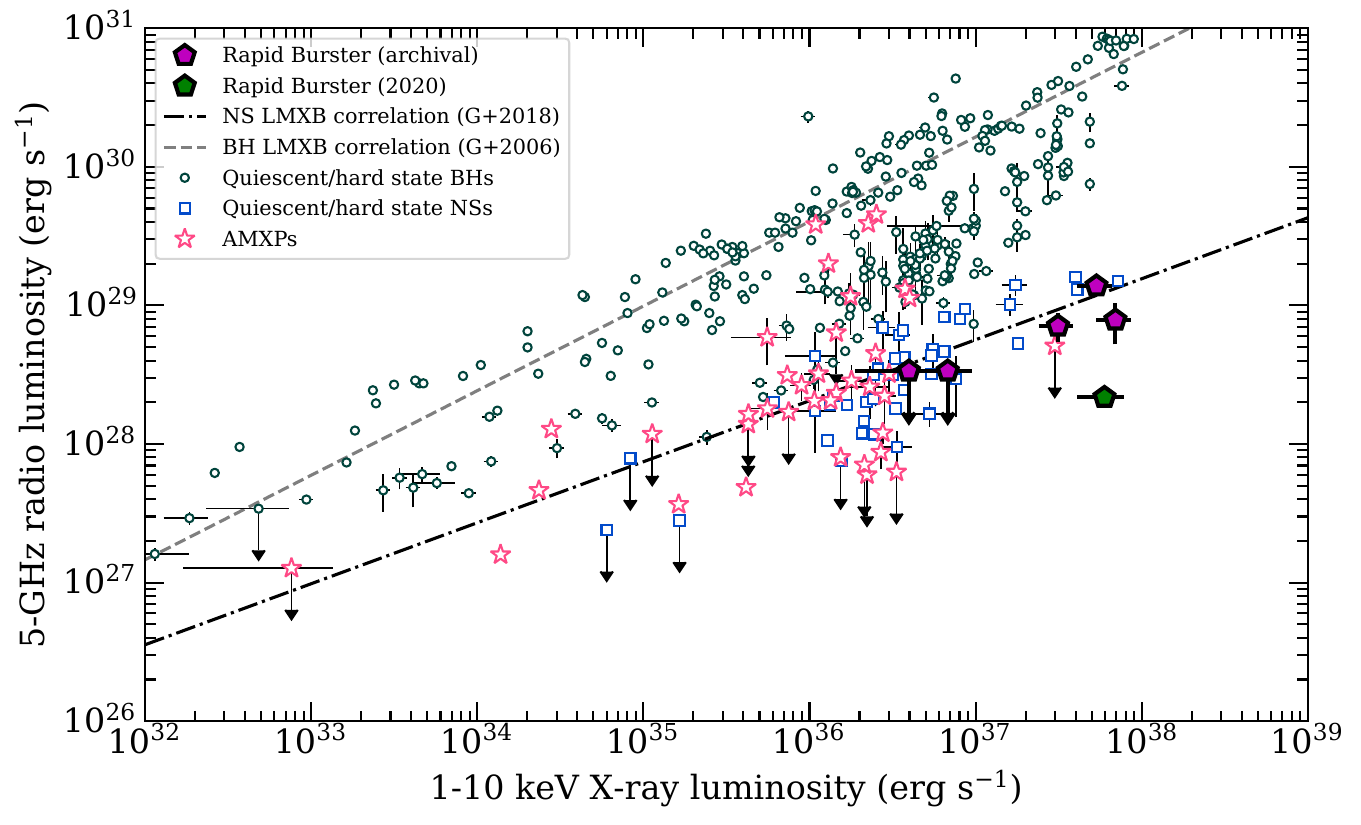}
 \caption{The X-ray -- radio luminosity plane for neutron star and black hole X-ray binaries. The open squares and open stars show hard-state atoll neutron stars and accreting millisecond X-ray pulsars, respectively; the open circles show black holes. The Rapid Burster, based on archival and 2020 data, is shown as the filled pentagons. The dashed and dash-dotted lines show the black hole and neutron star correlations as determined by \citet{gallo2006} and \citet{gallo18} Strikingly, the 2020 Rapid Burster observation (during a prolific bursting state), shows the source at its faintest detected radio luminosity, while its X-ray luminosity is similar to the radio-brighter epochs presented in \citet[][when no bursting was observed]{moore2000}. The comparison samples of black holes and neutron stars were collated by \citet{bahramian2022}.}
 \label{fig:LxLr}
\end{figure*}

In Figure \ref{fig:LxLr}, we show the X-ray -- radio luminosity diagram for LMXBs, based on the database by \citet{bahramian2022}. Capturing the Rapid Burster's observation-averaged behaviour, we plot the archival points, adapted from \citet{moore2000} and listed in Table \ref{tab:archival}, as the five purple pentagons \citep[due to the lack of X-ray information, we do not include the data from][]{tudor2022}. The 2020 campaign is plotted as the green pentagon. The three archival radio detections are consistent with the X-ray brightest hard state neutron star LMXBs; the two archival upper limits are only consistent with the radio-faintest neutron star LMXBs in the range between $L_X = 5\times10^{36}$ and $10^{37}$ erg/s. Interestingly, the Rapid Burster was significantly radio fainter, by a factor $\sim 5$, during the 2020 epoch than during the archival radio monitoring, despite a similar average X-ray luminosity.

\begin{figure*}
\includegraphics[width=\textwidth]{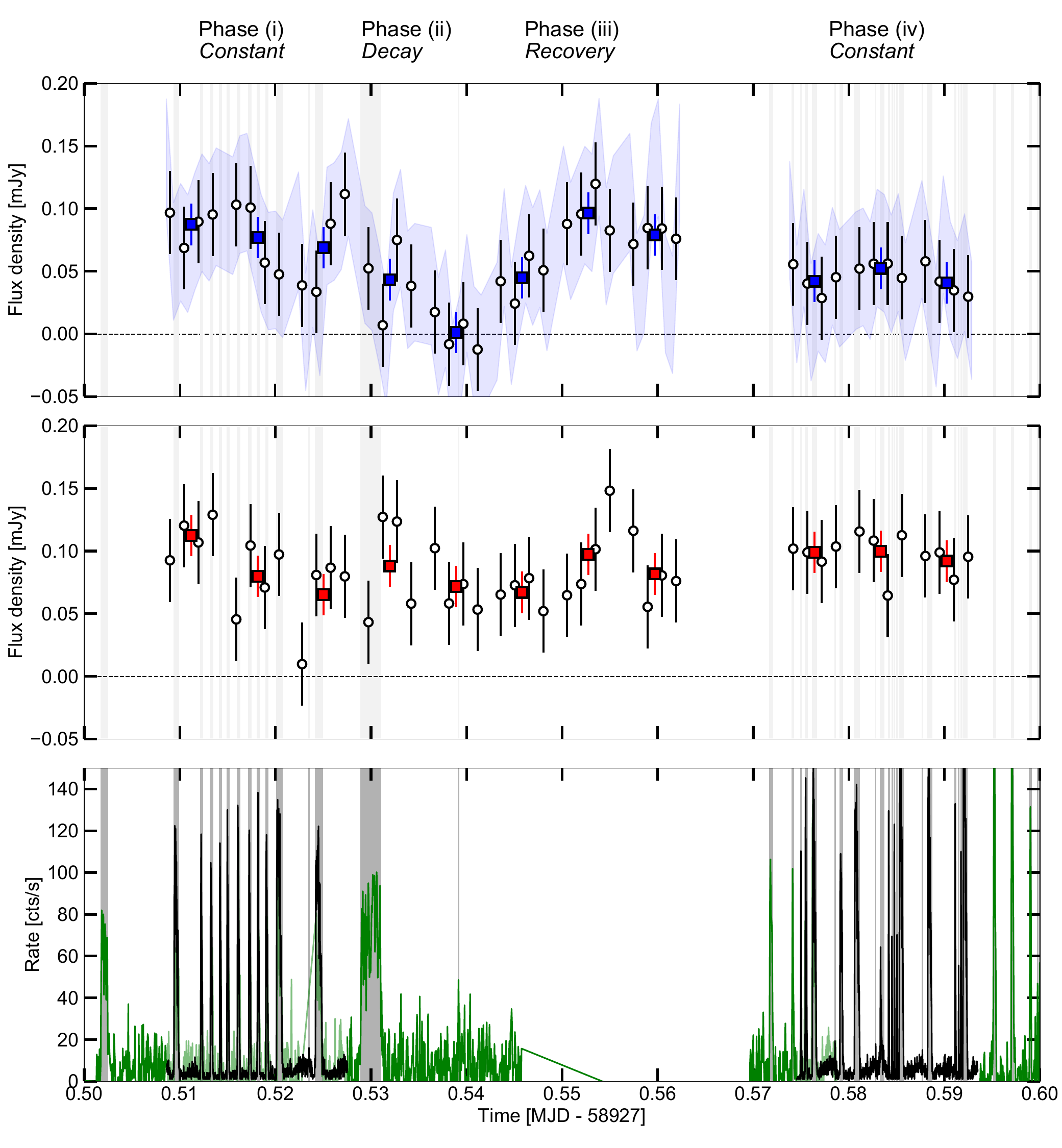}
 \caption{Radio and X-ray light curves of the Rapid Burster and the check source from the coordinated 2020 campaign. The top panel shows the radio light curve of the Rapid Burster, at three time resolutions: one and four measurements per scan (filled squares and open circles, respectively), and eight measurements per scan (the shaded band, indicating the region with the $1\sigma$ flux density uncertainties). The middle panel shows the radio light curve, at the former two time resolutions (following the same marker convention), for the check source. All radio flux densities shown in both panels are measured in the uv-plane. The bottom panel shows the X-ray light curves measured by \textit{Swift}/XRT in black (1 second time resolution) and \textit{INTEGRAL}/JEM-X {\color{white}in green} (5 second time resolution), where the latter have been multiplied by $1.5$ to better match the vertical extent of the curves. We identify Type-II X-ray bursts in the bottom panel as instances where the X-ray count rate exceeds 15 and 30 counts/second, for \textit{Swift} and \textit{INTEGRAL}, respectively. Those times are indicated as the shaded vertical bands in all panels. Above the top panel, the four qualitative phases of the Rapid Burster's radio light curve are indicated, as introduced in Section \ref{sec:phases}. }
 \label{fig:all_lcs}
\end{figure*}

\begin{figure}
\includegraphics[width=\columnwidth]{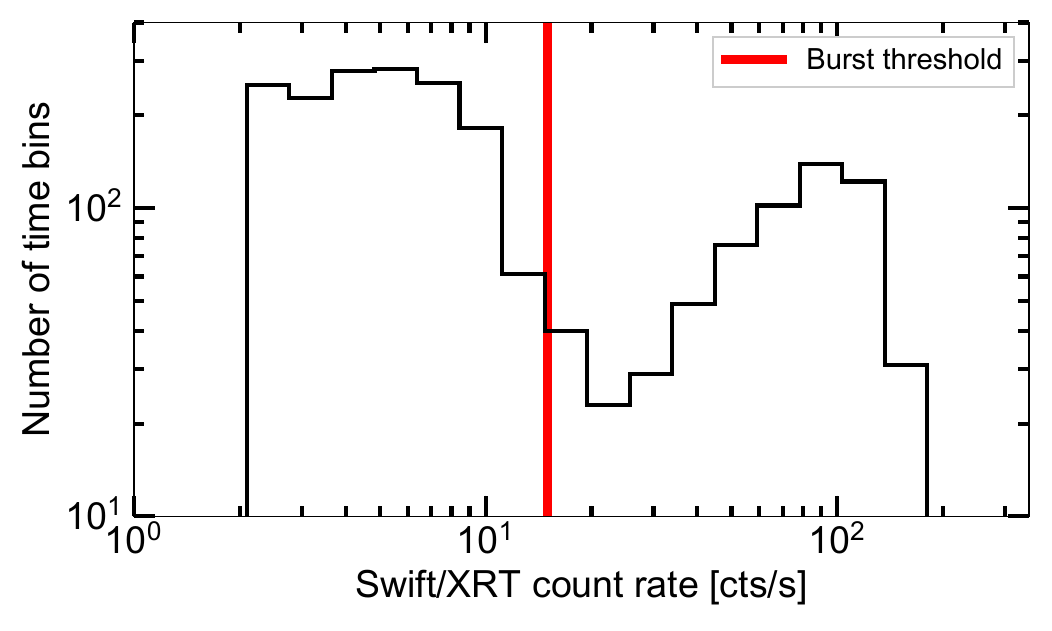}
\caption{A logarithmically-binned histogram of the \textit{Swift}/XRT count rates, showing two clear peaks due to burst and non-burst emission. The red line indicates the burst-rate-threshold, above which the bursts first start to contribute and then dominate the count rates.}
 \label{fig:rate_hist}
\end{figure}

\subsection{Time-resolved analysis}
\label{sec:results_resolved}

\begin{figure*}
\includegraphics[width=1.05\textwidth]{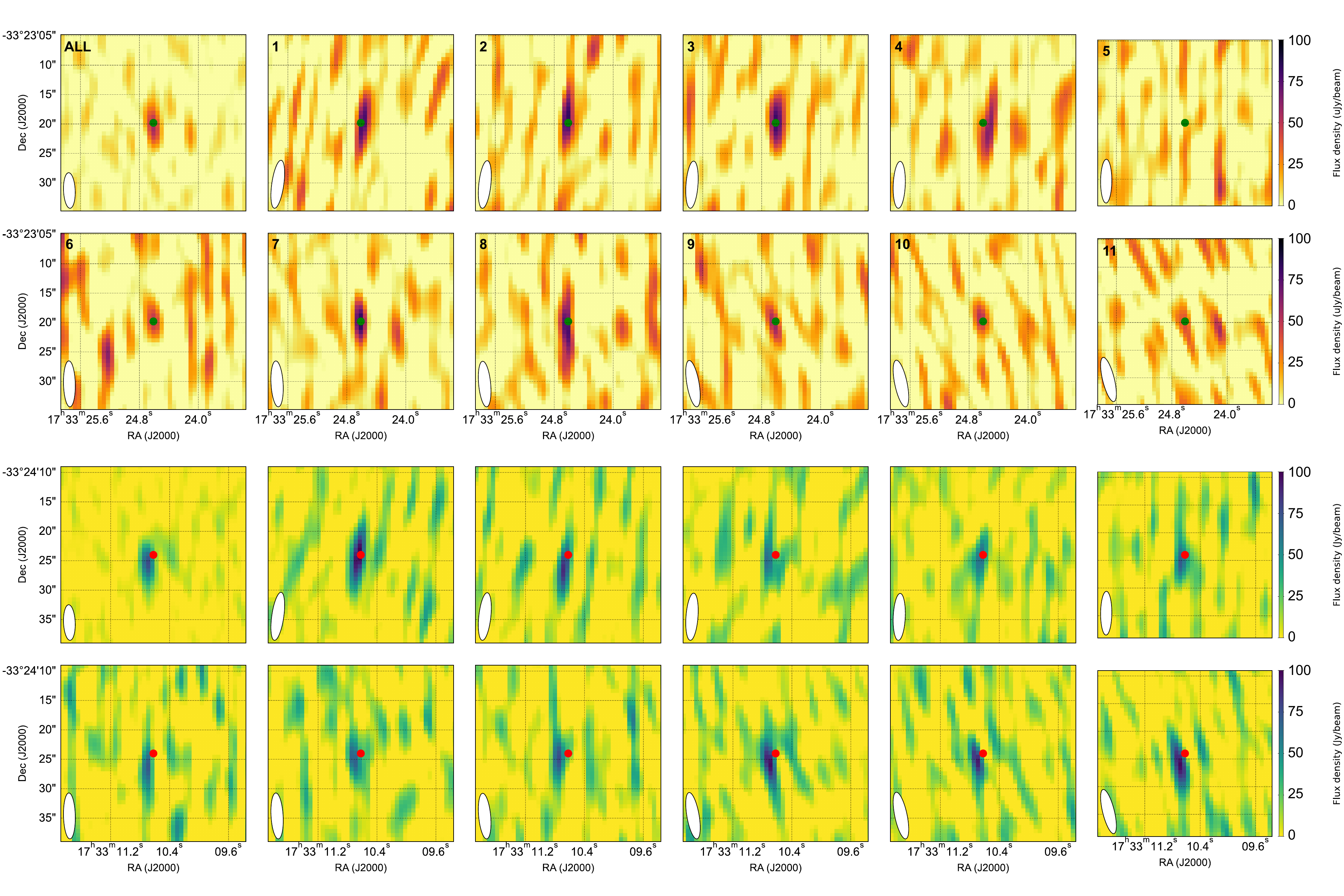}
 \caption{The inner $0.5\times0.5$ arcminute field of view around the Rapid Burster (top two rows) and check source (bottom two rows). The top left panel for both sources shows the time averaged image; the remaining eleven panels per source show the image for each of the scans of the target field (numbers for the top two rows as well). Both sources and all panels are shown with the same extent and scaling of their colormaps. The panels show that the Rapid Burster varies between $\sim 100$ $\mu$Jy and non-detection (scans 5, top right panel), as is similarly seen in the uv-plane measurements underlying the light curve in Figure \ref{fig:all_lcs}. Similarly, the bottom two rows show how the check source is always detected on top of a varying background. In all panels, the synthesized beam is shown in the bottom left: it can, as expected, be seen to rotate throughout the observation. The position of the source is shown with the point in each panel.}
 \label{fig:all_images}
\end{figure*}

\begin{figure}
\includegraphics[width=\columnwidth]{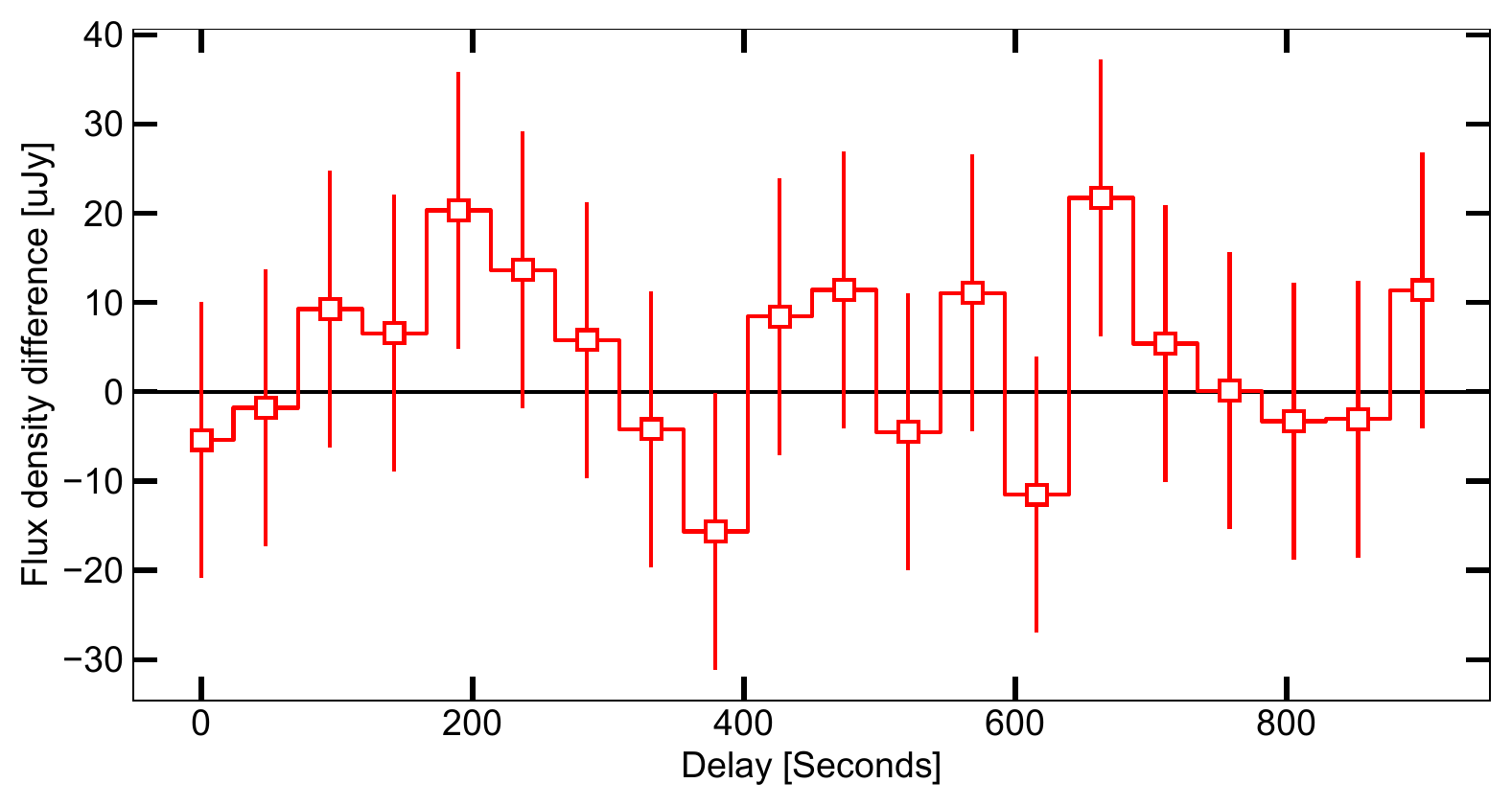}
 \caption{The difference between the flux density of the Rapid Burster measured during all burst times and all non-burst times, plotted as function of delay. The delay is defined as the shift in burst times used when measuring the combined burst and non-burst flux densities. The measured flux density differences are consistent with statistical variations around zero.}
 \label{fig:delay}
\end{figure}

During the coordinated 2020 observing campaign, the Rapid Burster resided in a state of prolific bursting. In this Section, we explore the time-resolved properties of the X-ray and radio observations. In Figure \ref{fig:all_lcs}, we show the radio and X-ray light curves of the Rapid Burster and the radio check source. The upper panel shows the Rapid Burster's radio flux density as a function of time at three time resolutions: one, four, and eight measurements per target scan, plotted as the squares, and open circles, and shaded band (indicating the range included within the $1\sigma$ uncertainties), respectively. The middle panel shows the flux densities at for the same two resolutions, but for the check source. Finally, the bottom panel shows the \textit{Swift}/XRT (black) and \textit{INTEGRAL}/JEM-X (green) X-ray light curves. The small gaps in the two radio light curves indicate secondary calibrator scans, while the larger gap corresponds to the primary calibrator scan; the latter was coordinated to overlap with the Earth occultation in the \textit{Swift} observation, visible as the gap in the bottom panel. 

In the X-ray light curve, a large number of Type-II bursts are evident. The \textit{Swift}/XRT light curve shows that, when the burst recurrence time allows, the intra-burst rate slowly increases after the end of the previous burst, before a dip preceeds the next bursts. Such behaviour, for instance seen after the Sun constraint window, is consistent with the pre- and post-burst dips known in both the Rapid Burster and Bursting Pulsar \citep{bagnoli2015,younes2015,court2018}. No Type-I bursts, with a characteristic exponential cooling tail, are observed. Due to their brightness, the Type-II bursts are easily identified by eye. After trialling several values, we find that a $15$ cts/s and $30$ cts/s threshold for \textit{Swift} and \textit{INTEGRAL}, respectively, select all bursts without mistaking stochastic variability in the quiescent emission for bursts (the difference in driven by the larger noise in the JEM-X light curve). We therefore adopt those values to define the burst start and end times, and indicate these times with light vertical bands in all panels. Most Type-II bursts occur with a relatively short recurrence time ($\sim 1$ minute) and duration (tens of seconds); the exceptions are the first plotted Type-II burst and the burst occurring around MJD 58927.53, both observed by \textit{INTEGRAL}: both are of longer duration and are followed by a comparatively long period without bursting. Using the \textit{Swift} burst threshold, we measure that $\sim 85$\% of the X-ray fluence in the observation arises during the times of Type-II bursts, which occur during $\sim 26$\% of the time. We plot a logarithmic histogram of \textit{Swift}/XRT fluxes in Figure \ref{fig:rate_hist} to further clarity.

The radio light curve of the Rapid Burster shows variability inconsistent with statistical variations around a constant: the $\chi^2_\nu$ compared to a constant mean flux density is $\chi^2_\nu = 2.7$ for the lowest time resolution (one measurement per 519-second scan). In comparison, the check source, at the same time resolution, is consistent with a constant at $\chi^2_\nu = 0.8$. At shorter time resolutions, the uncertainties on the Rapid Burster's flux density increase such that any apparent variability becomes consistent with a constant. The trends of these shorter duration measurements, however, follow the behaviour indicated by the longer time resolution light curve, further strengthening the argument that these variations at the lowest time resolution are physical and do not arise from statistical variations. Following \citet{vaughan2003}, we calculate the fractional variability in the Rapid Burster's lowest time resolution light curve as $F_{\rm var} = 38 \pm 5$\%. For the other time resolutions and the check source, we find $\chi^2_\nu < 1$, implying that the variance of the light curve is smaller than the mean square errors on the data. Therefore, this fractional variability is consistent with zero and cannot be calculated. 

The radio light curves in Figure \ref{fig:all_lcs} show the flux densities calculated through the uv-plane method described in Section \ref{sec:data_2020}. In particular, the Rapid Burster displays variations between $\sim 100$ $\mu$Jy to non-detections, characterised by a uv-plane flux density measurement consistent with zero. The first segment of the VLA observations, prior to the primary calibrator scan, in particular displays a striking evolution, first decreasing from this maximum to minimum radio level before recovering on a comparable time scale. To confirm that this evolution is not an artefact of the analysis method, we generate images of each individual scan, plotted for both the Rapid Burster (top half) and the check source (bottom half) in Figure \ref{fig:all_images}. The top left panel for both sources shows the time-averaged image. The check source is detected in each image, on top of a varying background, consistent with the uv-plane light curve in the middle panel of Figure \ref{fig:all_lcs}. The Rapid Burster, on the other hand, is indeed observed to change significantly throughout the observation, with a non-detection during the fifth target scan (top right panel). 

Linking the variability in the radio light curve to individual Type-II X-ray bursts is challenging for a host of reasons: firstly, the relatively low signal-to-noise ratio of the full observation, corresponding to a detection significance of $\sim 10\sigma$, causes any variability plotted for time scales shorter than individual scans, to be statistically insignificant. Furthermore, the majority of bursts last for less than 30 seconds. This short duration not only reduces the amount of radio data collected per burst, but also implies that the following non-burst period is shorter. Bursts that follow each other on time scales shorter than the characteristic time scale of any potential jet response, may not individually affect the jet but instead end up in a resonance, such that their effect may instead be collective. As a test to overcome this limitation of individual bursts, we fit the flux density of the Rapid Burster in the uv-plane using only the combined burst times or the combined non-burst times, as identified by the grey shaded regions in Figure \ref{fig:all_lcs}. We do not find a statistically significant difference between the collective burst and non-burst intervals in the radio flux density of the Rapid Burster:  $54\pm15$ $\mu$Jy and $59\pm6$ $\mu$Jy, respectively. As the radio emission originates from far down the jet, a delay is expected between any radio jet response to the X-ray behaviour. However, systematically shifting the start and end times of the burst and non-burst intervals with an increasing delay, does not reveal a systematic difference at any delay time scale either: in Figure \ref{fig:delay}, we show the flux density difference as a function of delay, finding a curve that is consistent with statistical scatter around zero. Given the faintness of the Rapid Burster in comparison with the sensitivity in the cumulative burst and non-burst intervals, this negative result is not surprising. 

\label{sec:phases}
Turning, therefore, to a more qualitative assessment, the scan-by-scan radio variability in Figure \ref{fig:all_lcs} can be summarized, broadly, in four (slightly overlapping/continuous) phases: (i) a constant, relatively bright phase around the source's maximum flux density in scans 1-3; (ii) a decay phase down to non-detection in scans 3-5; (iii) a recovery phase to maximum flux density in scans 6-7; and finally (iv) a constant phase consistent with the observation-averaged flux density in scans 9-11. As argued in the previous paragraph, linking the radio variability to individual X-ray bursts is challenging. However, it is worth exploring how these four radio phases relates to the overall bursting behaviour in X-rays. 

One possible interpretation of the radio variability is as follows: the two brightest radio phases -- phase (i) and the endpoint of the recovery in phase (iii) -- follow after the longest burst-free periods in the observation: the burst-free segments after the first \textit{INTEGRAL} burst and the {\color{white} segment} after the most fluent detected burst, centred on MJD 58927.53. The decay in phase (ii), on the other hand, appears to follow from the transition from rapidly repeating, short bursts (MJD 58927.51 -- 58927.52) to stronger, more fluent, and more slowly recurring bursts (after MJD 58927.52); the minimum in the light curve, i.e. the non-detection, occurs after the strongest Type-II burst, before recovery sets in. Finally, phase (iv) coincides with a relatively stable bursting period, where the bursts recur with slightly longer wait times and duration than between MJD 58927.51 -- 58927.52. This potential interpretation could tentatively be summarized as follows: X-ray bursting tends to weaken the jet, with more fluent and slower recurring bursts having a stronger effect, while longer periods of non-bursting lead to the recovery of the jet to its maximum level. Such potential, tentative links would affect the radio emission with a delay, estimated visually from the light curves, of the order of $\sim 5\times10^{-3} - 10^{-2}$ MJD, or $\sim$ 7 to 14 minutes. 

An alternative scenario should be considered as well, however: the response time in the jet may be substantially shorter, as expected when driven by the expected escape velocity from the neutron star \citep[e.g., $\sim 3$ minutes, as seen in response to Type-I bursts in 4U 1728-34 by][]{russell2024}. For such more immediate response, we observe that the radio brightest phases mentioned above occur during prolific X-ray bursting (Phase (i)), or during a period lacking X-ray coverage. The radio-faintest segment, during the fifth target scan, instead coincides with a lack of X-ray bursting. This alternative interpretation fits with a scenario where the jet is weakened \textit{in between} the bursts.

This ambiguity in interpretation is strengthened by the lack of radio observations during the aforementioned first long burst-free period (i.e. after the first \textit{INTEGRAL} burst) and the lack of both radio and X-ray information on the transition between phase (iii) and (iv). We also note that, due to the non-significance of the variability on shorter time scales, we do not attempt to extend these lines of qualitative arguments to this faster variability (or lack thereof). Therefore, the above scenarios should be assessed critically in the light of all (including archival) information on the Rapid Burster, which we will do in Section \ref{sec:discussion}. 

%\end{landscape}

\section{Discussion}
\label{sec:discussion}
    
While the unique X-ray bursting properties of the Rapid Burster have been known for decades, their effect on the binary's jet launching properties has remained elusive. The main results of our 2020 investigation of this question, using simultaneous VLA, \textit{Swift}, and \textit{INTEGRAL} observations, can be summarized as follows:
\begin{enumerate}
    \item The Rapid Burster is detected at a time averaged 0.5--10 keV X-ray flux of $(8.0\pm2.5)\times10^{-9}$ erg/s/cm$^2$ and radio flux density of $58.1 \pm 5.5$ $\mu$Jy; the latter is the faintest radio detection of the Rapid Burster to date. 
    \item The 2020 campaign coincides with a phase of prolific Type-II X-ray bursting, with \textit{Swift} detecting $85$\% of its total X-ray fluence during burst times.
    \item The radio luminosity of the Rapid Burster was significantly lower in 2020 than in the observations presented in \citet{moore2000} at a comparable time-averaged X-ray luminosity; interestingly, no Type-II bursts occurred during those archival, radio-brighter observations. 
    \item The radio flux density of the Rapid Burster shows significant variability between target scans, with a fractional variability of $F_{\rm var} = 38 \pm 5$\%, evolving between $96\pm16$ $\mu$Jy to non-detection. On shorter time scales (i.e., within scans), no statistically-significant variations are found. 
\end{enumerate}

In this Section, we will first compare these observed properties of the Rapid Burster with other neutron star LMXBs, focusing particularly on its comparative radio faintness and radio variability. Secondly, we will discuss the implications of our results and their broader context on the origin of Type-II bursts and on jet launching; discussing the latter topic both for the Rapid Burster specifically and in neutron star LMXBs more generally. 

\subsection{The 2020 campaign in context: historical observations and other LMXBs}

As shown in the X-ray -- radio luminosity diagram, the 2020 campaign reveals an unexpectedly faint radio counterpart of the Rapid Burster. To assess this faintness more quantitatively (particularly comparing across X-ray luminosities), we calculate the relative position of the Rapid Burster and other neutron star LMXBs with respect to the neutron star the X-ray -- radio luminosity correlation found by \citet{gallo18}:

\begin{equation}
\log L_{\rm R} -  \log L_{\rm R,0} = \alpha + \beta \left(\log L_{\rm X} - \log L_{\rm X,0}\right)  \text{ ,}
\end{equation}

\noindent where we do not include the scatter parameter, as we intend to quantify this scatter. We adopt the parameters from \citet{gallo18}, who find $\log L_{\rm R,0} = 28.57$, $\log L_{\rm X,0} = 36.20$, $\alpha=-0.17$ and $\beta=0.44$. Introducing renormalized luminosities $\overline{L_{\rm i}} = L_{\rm i}/ L_{\rm i,0}$, the above Equation is equivalent to

\begin{equation}
\log \left(\frac{\overline{L_{\rm R}}}{\overline{L_{\rm X}}^{\beta}}\right) = \alpha \text{.}
\end{equation}

\noindent Calculating the left hand side of this equation for all individual observations of neutron star LMXBs, we can assess their deviation from the best-fit correlation with respect to the overall correlation’s logarithmic normalization $\alpha$. 

\begin{figure*}
\includegraphics[width=\textwidth]{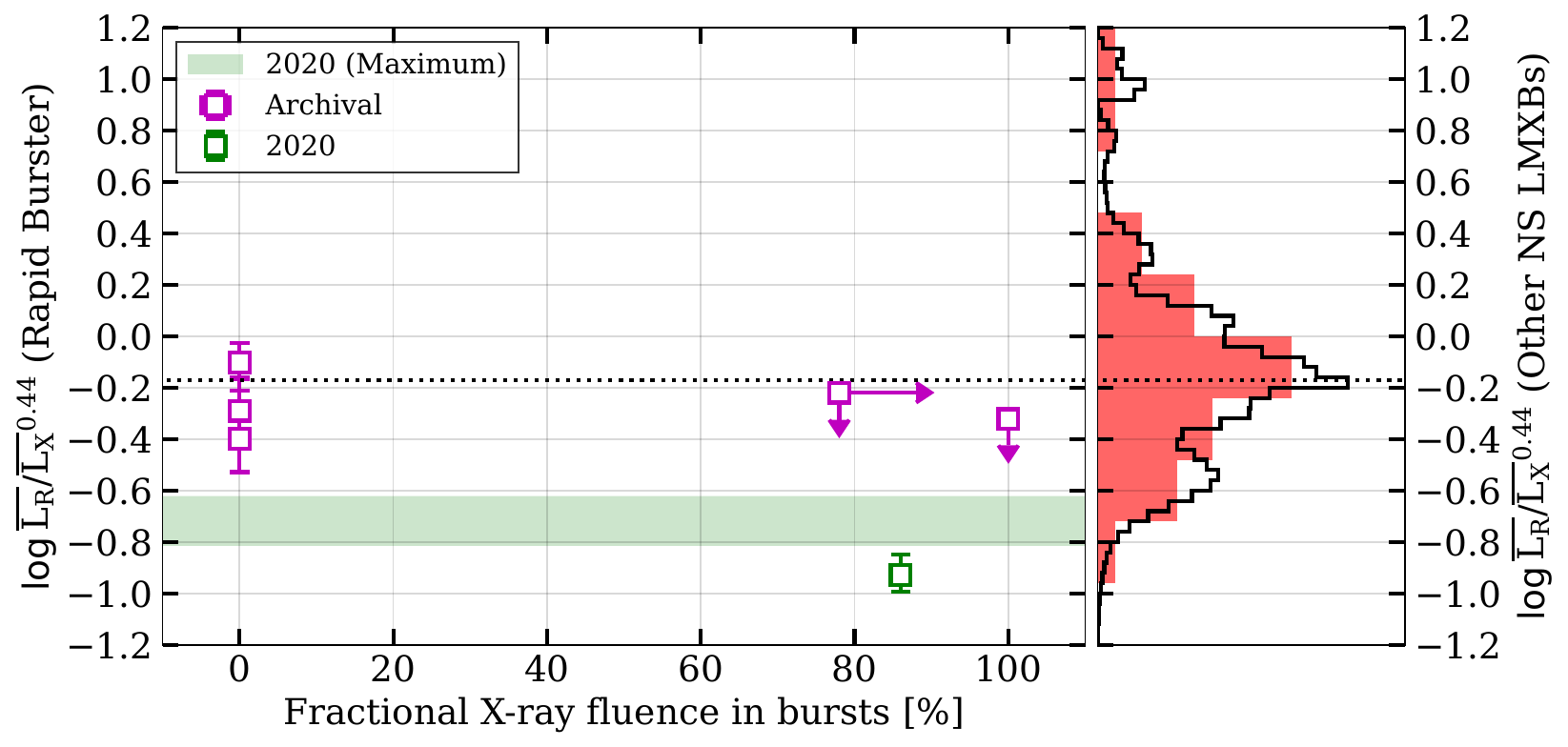}
 \caption{\textit{Left panel:} the relative radio brightness of the Rapid Burster as a function the fractional X-ray fluence detected during the times of Type-II bursts. The relative radio brightness is plotted on a logarithmic scale and defined as the ratio between the normalized radio luminosity $\overline{L_{\rm R}}$ and the normalized X-ray luminosity $\overline{L_{\rm X}}^{0.44}$, which captures the difference between the radio luminosity and the best-fit $L_{\rm X}$--$L_{\rm R}$ relation for neutron star LMXBs as measured by \citet{gallo18} and shown in Figure \ref{fig:LxLr} (see main text for details and normalization luminosities). The six data points show the observation-averaged data from the five archival observations from \citet{moore2000} and the 2022 campaign; the shaded horizontal band shows the maximum radio luminosity observed in the per-scan light curve of the Rapid Burster in 2020. In the latter, the fractional X-ray fluence in Type-II bursts is ill-defined, as we calculate this for full observations. \textit{Right panel:} a vertical histogram of the same relative radio brightness, calculated using the X-ray and radio luminosities of the comparison sample of neutron star LMXBs from \citet{bahramian2022}. The filled histogram shows this comparison sample; the unfilled histogram shows the same data after including their uncertainties on X-ray and radio luminosity via a Monte-Carlo approach. \textit{In both panels}, this dotted line shows the normalization value measured by \citet{gallo18} for neutron star LMXBs, unsurprisingly matching the histogram peak.}
 \label{fig:LxLrRatio}
\end{figure*}

In Figure \ref{fig:LxLrRatio}, we plot this quantity for the Rapid Burster, as a function of the X-ray fluence observed during times of Type-II bursts (left panel). We also show, using a rotated histogram, all other neutron star LMXBs in the comparison sample in Figure \ref{fig:LxLr} (right panel). For this latter histogram, we show two realisations: a histogram calculated directly from the observed luminosities, and one calculated from $10^4$ re-constructions of the comparison sample by varying each luminosity within its uncertainty (assuming Gaussian luminosity errors and a 10\% error for luminosities originally reported without uncertainties). As expected, the peak of this distribution is consistent with $\alpha = -0.17$ (dotted line), as found by \citet{gallo18}. 

For the 2020 Rapid Burster observation, we directly measured the fluence during times of Type-II bursts from the \textit{Swift} light curve; for the archival data, we used the information reported in \citet{moore2000} regarding the \textit{RXTE}/PCA light curves: the number of Type-II bursts, their average duration, and the average count rates during the entire observation and during the bursts. Combining this with the exposure of each observation and assuming that the bursts are achromatic (implying the same rate to flux conversion during and in between bursts), we calculated the fractional fluence during burst times. 

During the 2020 observation, the Rapid Burster showed a deviation of $\log \left(\overline{L_{\rm R}} / \overline{L_{\rm X}}^{\beta}\right) = -0.93^{+0.07}_{-0.08}$, amongst the radio faintest deviations seen in the larger neutron star LMXB sample. The two archival upper limits of the Rapid Burster, instead, lie closer to the mean $\alpha$, as do the three measurements from the archival radio detections. We therefore conclude that the 2020 observation of the Rapid Burster was radio faint compared to (i) archival studies of the same source without Type-II bursts; and (ii) to the full neutron star LMXB sample, when correcting for the sample's dependence on X-ray luminosity.

Before turning to the Rapid Burster's variability, we should consider several important caveats to the above analysis. Firstly, neutron star LMXB show significant scatter around their best-fit relation, whose slope may be affected by a bias against publication of radio upper limits, particularly at low X-ray luminosities. Therefore, the correction of the X-ray luminosity dependence may be similarly biased by assuming $\beta = 0.44$. Furthermore, as \citet{gallo18} explore, different sub-classes of neutron star LMXB may show different accretion flow -- jet couplings. Finally, the histograms plotted in Figure \ref{fig:LxLrRatio} are based on radio-detected neutron stars LMXBs. The low-$\alpha$ tail may therefore be artificially over-pronounced as the number low-$\alpha$ detections can be be limited by radio sensitivity. A striking example of this behaviour is given by Aql X-1, for which \citep{gusinskaia2020} present a rapid decrease of its radio luminosity during the decay of its 2016 outburst. The three radio non-detections of Aql X-1 during this decay show a strong suppression compared to the correlation we applied in our analysis, with upper limits on $\alpha$ between $-0.9$ and $-1.0$. These values, similar to the Rapid Burster 2020 campaign, highlight these sensitivity-limited biases in our approach. While these effects do not counter the conclusion that the Rapid Burster was radio faint during the 2020 campaign, they should be kept in mind when interpreting the exact value of $\alpha$. 

While radio faint, the Rapid Burster showed remarkable radio variability of its radio jet on the time scale of individual target scans. While rarely observed (or investigated; see Section \ref{sec:future} and below) amongst atoll sources, radio variability is regularly seen in other (classes of) neutron stars in LMXBs and similar systems. Z sources, for instance, have long been known to vary strongly in radio properties, changing between different branches of their Z track \citep{hjellming1990a,hjellming1990b,migliari06} and potentially launching transient ultra-relativistic outflows \citep{fomalont2001a,fomalont2001b,motta2019}. Similar levels of fractional radio variability to the Rapid Burster were observed in Swift J1858.6-0814 \citep{vandeneijnden2020}, which \citet{vincentelli2023} modelled as arising from jet ejecta following instabilities in the inner accretion flow. Recently, \citet{russell2024} reported the discovery of radio jet brightenings directly succeeding thermonuclear (Type-I) bursts in 4U 1728-34. During X-ray-dominated states of (candidate) transitional milli-second pulsars (tMSPs), significant and even structured radio variability has been observed: for instance, the prototypical tMSP PSR J1023+0038 shows both radio flaring and radio moding, sharply transitioning between low and high states which are strongly anti-correlated with similarly-sharp X-ray moding \citep{bogdanov2018}. The candidate tMSP CXOU J110926.4-650224 was more recently reported to show radio brightening throughout a coordinated X-ray and radio observation, potentially in response to the onset of X-ray flaring \citep{cotizelati2021}.  

The radio variability in the Rapid Burster therefore fits within a sub-class of accreting neutron stars: those that display a jet/outflow response, following sudden changes in the inner accretion flow that are triggered by instabilities in the flow itself  (Swift J1858.6-0814), by track changes of Z-sources \citep[][]{penninx1988,hjellming1990a,hjellming1990b}, by the influence of the neutron star's surface (thermonuclear bursts in 4U 1728-34), or by magnetic fields \citep[likely playing a central role in tMSPs; see e.g.,][for a recent review]{papitto2022}. It should be noted that the literature contains few detailed investigations of the radio variability on short time scales for atoll neutron stars and AMXPs that show no peculiar X-ray variability. In the atoll systems and AMXPs where such radio variability of their compact jet was investigated, no evidence for significant variability was detected within radio observation \citep{russell2018,gusinskaia2020b}. However, the subset of sources where such searches are performed and reported remains small, raising the question to what extent peculiar X-ray behaviour introduces a strong bias motivating radio variability searches. Yet, even within this context of other accretion neutron stars, the Rapid Burster and its radio variability remain unique: it remains the only radio-detected source with Type-II source, allowing for a further comparison with expectations from Type-II burst and neutron star jet models. 

\subsection{Implications for Type-II bursts and neutron star jets}

The relative position of the Rapid Burster in the X-ray -- radio luminosity diagram may be understood as either relatively X-ray bright or relatively radio faint, or a combination of both. The X-ray bright scenario may arise if the radio jet is coupled to the accretion flow in between Type-II bursts, but does not substantially respond to the increased mass accretion during the bursts. In other words, the jet is de-coupled from the accretion rate during bursts, and calculating a mean X-ray luminosity during the observation over-estimates the accretion rate driving the jet. In the radio-faint scenario, on the other hand, the presence of the Type-II bursts would weaken the jet either during or in between bursts. As a result, while the a-chromatic nature of Type-II bursts implies that the average X-ray luminosity traces the average mass accretion rate, the resulting jet radio luminosity is weaker. 

Our observational results argue that the first scenario is unlikely. Scaling the mean non-burst \textit{Swift} rate, the non-burst X-ray luminosity of the Rapid Burster is $(1.2\pm0.4)\times10^{37}$ erg/s. Even at that level, the observed radio luminosity is substantially fainter than the mean neutron star X-ray -- radio luminosity correlation at that non-bursting X-ray luminosity. This remains in contrast with the archival, non-bursting observations, where the observed radio luminosity is consistent with the mean neutron star LMXB correlation (see also Figure \ref{fig:LxLrRatio}). On the other hand, the strength of this argument is limited by the range that individual neutron star LMXBs show in the inferred slopes of their X-ray -- radio luminosity correlations: a correlation for the Rapid Burster steeper than the best-fitting slope of the full sample would counter this argument. However, the X-ray bright scenario does also not account for the observed radio variability: in this scenario, the jet is effectively de-coupled from the bursts, and may potentially not be expected to show the significant levels of variability observed in the 2020 campaign. As discussed earlier, a larger number of radio variability measurements in atoll sources would provide a crucial further test of this variability argument, by fully addressing the uniqueness of such variability amongst neutron star LMXBs. 

Alternatively, the radio-faint scenario, where the jet weakens as a result of the presence of Type-II bursts, can account for both the faintness and variability. This scenario can fit the two possible interpretations of the four phases introduces in the radio light curve in Section \ref{sec:phases}: either the jet is weakened by bursts, and the fading and re-brightening of the jet respectively follow an intense phase of bursting, including the most powerful burst observed in the campaign and the longest burst-free period; alternatively, the jet is weakened in between bursts, as the radio-faintest segment overlaps with a burst-less period. In principle, the radio faintness and variability may be explained by both scenarios: these two observables reveal that the jet is not constantly launched at a constant and high luminosity, but do not necessarily constrain whether the jet is correlated or anti-correlated with the Type-II bursts. Therefore, causality comes into play, through the delay that is physically expected, and routinely observed in any radio response to X-ray variability \citep{atetarenko2019_cygx1,vincentelli2023,russell2024}. A longer delay implies that the causal burst-jet relation is more likely to be destructive (i.e., an anti-correlation); instead, associating the phase of radio non-detection with the coincident burst-free period, requires a shorter time for the jet response to reach the radio-emitting regions. We finally note that a third, but minor feature is present in the X-ray light curve: brief pre- and post-burst dips. Due to their relative weakness compared to the fluence changes between burst and non-burst phases, we deem it unlikely that these dips cause a significant effect in the radio light curve; testing the association explicitly is beyond our current data.

It is worth emphasizing that this potential effect of the presence of Type-II bursts -- weakening or even destroying the jet -- is opposite to the effect seen in most other LMXBs. Typically, increases in accretion rate and instabilities in the accretion flow are observed and modelled to lead to a more powerful compact jet or to the emergence of a second type of outflow \citep[transient ejecta, for instance, or expanding plasmoids suggested in tMSPs;][]{papitto2022}. As shown in Figure \ref{fig:LxLrRatio}, the maximum flux density reached by the system in the scan-by-scan radio light curve does not bridge the luminosity gap with the archival observations. This remaining faintness can be explained by the short recurrence time scale of the bursts, not allowing the jet to fully build up after or during successive bursts, depending {\color{white} on} the causal relation. These two properties of the observed radio response warrant a further discussion in the light of more specific jet and Type-II bursts models. So far, we have only associated the Type-II bursts with an increase in accretion rate, following model-independent empirical results. We will now briefly discuss our results specifically in the context of magnetic gating models for the bursts.

The magnetic gating model for Type-II bursts \citep{spruit1993,dangelo2010} poses that the bursts are the result of a centrifugal barrier created by the neutron star's magnetosphere: in between bursts, the magnetospheric radius where the accretion flow is truncated, is located slightly outside the co-rotation radius. The resulting centrifugal barrier leads to a gradual build-up of transferred material trapped in the inner disk, resulting in a low overall accretion rate and an increasing ram pressure from the flow. The latter pushes the magnetospheric radius inwards, up until the point where it crossed the co-rotation radius and the barrier vanishes, resulting in a burst wherein the build-up reservoir depletes and the system returns to its original configuration. Empirically, this scenario is supported by the observed correlation between recurrence time and burst fluence \citep{bagnoli2015}; moreover, X-ray reflection spectroscopic studies have revealed a significant truncation of the inner accretion flow in between the Type-II bursts of both the Rapid Burster and the Bursting Pulsar \citep{degenaar2014,vandeneijnden2017}. On the other hand, the uniqueness of the Rapid Burster and Bursting Pulsar within the much larger LMXB population remains poorly understood and cannot be straightforwardly explained by the gating model. Still, the gating model arguably remains the most well-developed and empirically-tested Type-II burst model in the literature. 

Combining this magnetic gating mechanism with the first potential scenario -- the tentative inference that the jet is fainter during bursts -- we can envision a scenario where the jet is launched from regions close to or beyond the magnetospheric radius, as the mass reservoir builds up. Given the large observed disk truncation in both the Rapid Burster \citep[$41.8^{+6.7}_{-5.3}$ $R_g$;][]{vandeneijnden2017} and the Bursting Pulsar \citep[$85.0\pm10.9$ $R_g$;][]{degenaar2014}, that scenario implies the jet is not launched from the inner tens of gravitational radii around the neutron star. Such a launch radius appears at odds with its expected location in the classical \citet{blandford1982} jet model, where toroidal magnetic field lines threading the innermost accretion flow are spun up by the flow's differential rotation. Instead, the central role for the neutron star magnetic field in controlling the accretion flow dynamics is consistent with models where the jet is powered by the spin of the neutron star, as its magnetic field lines are opened up by the inflowing material \citet{parfrey2016}. In such models, as more recently also demonstrated in computational simulations \citep{parfrey2023,das2022,das2024,berthier2024}, material is launched from outside the magnetospheric radius, regardless of its exact location. Moreover, when discussing jet quenching, \citet{parfrey2016} discuss how quenching is expected once the accretion rate becomes sufficiently high and the disk reaches the neutron star surface, such that the entire magnetosphere is dominated by accreting matter and its ability to launch an outflow is thereby suppressed. During Type-II bursts, the accretion rate increase may then indeed cause the jet to quench. In other words, such magneto-rotational neutron star jet models may explain both the launching of the jet in between bursts, and their potentially detected decay caused by bursts. We note that these models operate differently than propeller-type scenarios, where the rotating magnetosphere expels mass from the binary \citep{illarionov1975}: a mass reservoir should instead gradually build up beyond the magnetospheric radius in between the bursts.

In the alternative scenario, where the jet responds faster and is weaker in between Type-II bursts, the above reasoning leads to a different conclusion. In that case, a stronger jet is observed when material is rapidly accreting within the co-rotation radius, therefore arguing instead for \citep{blandford1982} type jet models. The limited time scale of the bursts may explain the relative faintness of the jet. In this scenario, a more accurate spectral index measurement, constraining the jet type, is of particular interest: does the swiftness of the accretion-driven burst lead to the launch of transient ejecta with a steep spectrum? Additionally, this scenario may suggest a different jet-launch scenario compared to more strongly-magnetized neutron stars, where magneto-rotational models likely operate \citep{vandeneijnden2021}.

If future observations confirm the potential jet response to Type-II bursts, we cannot directly conclude that this same jet model underlies all neutron star LMXB jet launching. Importantly, the relatively strong magnetic field, truncating and gating the accretion flow, does not play the same central role in all accreting neutron star systems. Without the strong disk truncation, which is rarely observed to the extent of the Rapid Burster and Bursting Pulsar \citep{ludlam2024}, the above arguments concerning the \citet{blandford1982} type models disappear. Jet launching in accreting neutron stars may instead be caused by a mixture of these various mechanisms, depending on the dynamical influence of the magnetic field; even in the Rapid Burster itself, when no bursts occur (i.e., at high accretion rate), different jet launch models may be dominant. 

Understanding whether all neutron star jets are launched through similar models requires a more detailed understanding of the jet properties themselves: moving beyond their mere detection and luminosity evolution to constrain outflow speeds, geometries, and powers. The recent study by \citet{russell2024} reports the response of a radio jet to Type-I bursts, which opens up a new method to measure such properties systematically in a large number of Type-I bursters. In our 2020 observations, the Rapid Burster does not reach sufficient signal-to-noise to obtain similar measurements, nor to allow us to break up the VLA observations in different frequency bands and track changes as they propagate down the jet. While the order-of-magnitude time scales of the two potential burst-jet associations discussed in Section \ref{sec:phases} ($7$--$14$ minutes versus significantly shorter, for a jet weakened \textit{during} or in \textit{between bursts}, respectively) are similar to the response time scale in 4U 1728-34 and Swift J1858.6-0814, we cannot perform the detailed modeling that subsequently constrains the jet's properties: such modelling requires an accurate delay measurement and multi-band radio monitoring. 

\subsection{Future outlook}
\label{sec:future}

To further investigate these potential burst and jet model implications, we envision three types of future observing campaigns. Firstly, the majority of published measurements or mentions of radio variability within observations are motivated by peculiar X-ray behaviour: bursting of different types, flaring, or moding. Therefore, we currently lack a comparison sample characterizing the radio variability in ordinary atoll neutron stars. Without such a base-level comparison, we are limited in assessing whether and to what extent the Rapid Burster’s radio variability is unique. Constructing such a comparison sample does not require new observations, but instead a dedicated and systematic re-analysis of published VLA observations of atolls at higher time resolution.

Secondly, the low signal-to-noise during the 2020 campaign may be countered by a longer coordinated observation during a phase dominated by longer and more slowly recurring Type-II bursts. If a causal relation indeed exists between the X-ray fluence during burst times and the suppression of the jet luminosity, it may not be possible to catch the source at higher radio luminosity while bursting -- in that case, such a longer campaign would benefit substantially from the improved sensitivity of future instruments such as the Square Kilometre Array or the next-generation Very Large Array \citep{selina18}. However, the longer recurrence time between powerful Type-II bursts, typically seen more early in the outburst \citep{bagnoli2015}, would ease the detection of a causal association between individual bursts and the response of the jet -- distinguishing between a constructive or destructive effect of the bursts. As a subsequent step, performing such a campaign at multiple observing frequencies would allow for the first measurements of jet speed and geometry in the Rapid Burster. 

Finally, instead of searching for a causal relation between the jet and individual bursts, a full outburst monitoring campaign could instead focus on the evolution of other, indirect observables of the jet-burst connection: the X-ray and radio luminosity, the fractional radio variability, and the fluence during burst times, for instance. Tracking such properties in tandem across the outburst could further test whether Type-II bursts suppress the jet. Such observations would require less strict simultaneity between X-ray and radio, where the maximum allowable time difference would be driven by the typical time scales of significant variations in X-ray or radio luminosity, and in bursting rates and properties, across the outburst.

\section{Acknowledgements}
For the purpose of open access, the authors have applied a Creative Commons Attribution (CC-BY) licence to any Author Accepted Manuscript version arising from this submission. The authors want to express their gratitude to the VLA, \textit{Swift}, and \textit{INTEGRAL} schedulers in their work to set up and perform the coordinated observations underlying this work, particularly during the unprecedented circumstances of the then-emerging COVID-19 pandemic. We thank the \textit{INTEGRAL} PI for approving the DDT observations used in this work. JvdE acknowledges a Warwick Astrophysics prize post-doctoral fellowship made possible thanks to a generous philanthropic donation. This research has made use of the MAXI data provided by RIKEN, JAXA, and the MAXI team. This research has made use of NASA's Astrophysics Data System Bibliographic Services. We acknowledge the use of public data from the \textit{Swift} data archive. This research was partly based on observations with INTEGRAL, and ESA project with instruments and science data centre funded by ESA member states (especially the PI countries: Denmark, France, Germany, Italy, Switzerland, Spain), and Poland, and with the participation of Russia and the USA. The National Radio Astronomy Observatory is a facility of the National Science Foundation operated under cooperative agreement by Associated Universities, Inc.

\section*{Data Availability}
A GitHub reproduction repository will be made public upon acceptance of this paper and will contain CASA analysis scripts, X-ray spectra and model files, X-ray and radio light curve files, as well as analysis and plotting scripts. An archived and stable release of this reproduction repository at the time of paper accceptance will also be made available on Zenodo {\color{white} via \url{https://github.com/jvandeneijnden/RapidBursterRadio}}. The unreduced VLA observations can be accessed publicly via the NRAO data archive (\url{data.nrao.edu}) under project code 20A-172. 

\bibliographystyle{mnras}
\bibliography{main.bib}

\begin{thebibliography}{}
\makeatletter
\relax
\def\mn@urlcharsother{\let\do\@makeother \do\$\do\&\do\#\do\^\do\_\do\%\do\~}
\def\mn@doi{\begingroup\mn@urlcharsother \@ifnextchar [ {\mn@doi@} {\mn@doi@[]}}
\def\mn@doi@[#1]#2{\def\@tempa{#1}\ifx\@tempa\@empty \href {http://dx.doi.org/#2} {doi:#2}\else \href {http://dx.doi.org/#2} {#1}\fi \endgroup}
\def\mn@eprint#1#2{\mn@eprint@#1:#2::\@nil}
\def\mn@eprint@arXiv#1{\href {http://arxiv.org/abs/#1} {{\tt arXiv:#1}}}
\def\mn@eprint@dblp#1{\href {http://dblp.uni-trier.de/rec/bibtex/#1.xml} {dblp:#1}}
\def\mn@eprint@#1:#2:#3:#4\@nil{\def\@tempa {#1}\def\@tempb {#2}\def\@tempc {#3}\ifx \@tempc \@empty \let \@tempc \@tempb \let \@tempb \@tempa \fi \ifx \@tempb \@empty \def\@tempb {arXiv}\fi \@ifundefined {mn@eprint@\@tempb}{\@tempb:\@tempc}{\expandafter \expandafter \csname mn@eprint@\@tempb\endcsname \expandafter{\@tempc}}}

\bibitem[\protect\citeauthoryear{{Arnaud}}{{Arnaud}}{1996}]{xspecref}
{Arnaud} K.~A.,  1996, in {Jacoby} G.~H.,  {Barnes} J.,  eds,  Astronomical Society of the Pacific Conference Series Vol. 101, Astronomical Data Analysis Software and Systems V. p.~17

\bibitem[\protect\citeauthoryear{{Avakyan}, {Neumann}, {Zainab}, {Doroshenko}, {Wilms}  \& {Santangelo}}{{Avakyan} et~al.}{2023}]{avakyan2023}
{Avakyan} A.,  {Neumann} M.,  {Zainab} A.,  {Doroshenko} V.,  {Wilms} J.,   {Santangelo} A.,  2023, \mn@doi [\aap] {10.1051/0004-6361/202346522}, \href {https://ui.adsabs.harvard.edu/abs/2023A&A...675A.199A} {675, A199}

\bibitem[\protect\citeauthoryear{{Bagnoli}, {in't Zand}, {D'Angelo}  \& {Galloway}}{{Bagnoli} et~al.}{2015}]{bagnoli2015}
{Bagnoli} T.,  {in't Zand} J.~J.~M.,  {D'Angelo} C.~R.,   {Galloway} D.~K.,  2015, \mn@doi [\mnras] {10.1093/mnras/stv330}, \href {https://ui.adsabs.harvard.edu/abs/2015MNRAS.449..268B} {449, 268}

\bibitem[\protect\citeauthoryear{{Bahramian} \& {Degenaar}}{{Bahramian} \& {Degenaar}}{2023}]{degenaarbahramian}
{Bahramian} A.,  {Degenaar} N.,  2023, in , Handbook of X-ray and Gamma-ray Astrophysics. Edited by Cosimo Bambi and Andrea Santangelo.
p.~120, \mn@doi{10.1007/978-981-16-4544-0_94-1}

\bibitem[\protect\citeauthoryear{{Bahramian} \& {Rushton}}{{Bahramian} \& {Rushton}}{2022}]{bahramian2022}
{Bahramian} A.,  {Rushton} A.,  2022, {bersavosh/XRB-LrLx\_pub: update 20220908}, \mn@doi{10.5281/zenodo.7059313}

\bibitem[\protect\citeauthoryear{{Blandford} \& {K{\"o}nigl}}{{Blandford} \& {K{\"o}nigl}}{1979}]{blandford1979}
{Blandford} R.~D.,  {K{\"o}nigl} A.,  1979, \mn@doi [\apj] {10.1086/157262}, \href {https://ui.adsabs.harvard.edu/abs/1979ApJ...232...34B} {232, 34}

\bibitem[\protect\citeauthoryear{{Blandford} \& {Payne}}{{Blandford} \& {Payne}}{1982}]{blandford1982}
{Blandford} R.~D.,  {Payne} D.~G.,  1982, \mn@doi [\mnras] {10.1093/mnras/199.4.883}, \href {https://ui.adsabs.harvard.edu/abs/1982MNRAS.199..883B} {199, 883}

\bibitem[\protect\citeauthoryear{{Blandford}, {Meier}  \& {Readhead}}{{Blandford} et~al.}{2019}]{blandford2019}
{Blandford} R.,  {Meier} D.,   {Readhead} A.,  2019, \mn@doi [\araa] {10.1146/annurev-astro-081817-051948}, \href {https://ui.adsabs.harvard.edu/abs/2019ARA&A..57..467B} {57, 467}

\bibitem[\protect\citeauthoryear{{Bogdanov} et~al.,}{{Bogdanov} et~al.}{2018}]{bogdanov2018}
{Bogdanov} S.,  et~al., 2018, \mn@doi [\apj] {10.3847/1538-4357/aaaeb9}, \href {https://ui.adsabs.harvard.edu/abs/2018ApJ...856...54B} {856, 54}

\bibitem[\protect\citeauthoryear{{CASA Team} et~al.,}{{CASA Team} et~al.}{2022}]{casa2022}
{CASA Team} et~al., 2022, \mn@doi [\pasp] {10.1088/1538-3873/ac9642}, \href {https://ui.adsabs.harvard.edu/abs/2022PASP..134k4501C} {134, 114501}

\bibitem[\protect\citeauthoryear{{Calla}, {Barathy}, {Snagal}, {Bhandar}, {Deshpande}  \& {Vyas}}{{Calla} et~al.}{1980}]{calla1980}
{Calla} O.~P.~N.,  {Barathy} S.,  {Snagal} A.~K.,  {Bhandar} S.~M.,  {Deshpande} M.~R.,   {Vyas} H.~O.,  1980, \iaucirc, \href {https://ui.adsabs.harvard.edu/abs/1980IAUC.3458....1C} {3458, 1}

\bibitem[\protect\citeauthoryear{{Carotenuto} et~al.,}{{Carotenuto} et~al.}{2021}]{carotenuto2021}
{Carotenuto} F.,  et~al., 2021, \mn@doi [\mnras] {10.1093/mnras/stab864}, \href {https://ui.adsabs.harvard.edu/abs/2021MNRAS.504..444C} {504, 444}

\bibitem[\protect\citeauthoryear{{Combi} \& {Siegel}}{{Combi} \& {Siegel}}{2023}]{combi2023}
{Combi} L.,  {Siegel} D.~M.,  2023, \mn@doi [\prl] {10.1103/PhysRevLett.131.231402}, \href {https://ui.adsabs.harvard.edu/abs/2023PhRvL.131w1402C} {131, 231402}

\bibitem[\protect\citeauthoryear{{Corbel}, {Nowak}, {Fender}, {Tzioumis}  \& {Markoff}}{{Corbel} et~al.}{2003}]{corbel03}
{Corbel} S.,  {Nowak} M.~A.,  {Fender} R.~P.,  {Tzioumis} A.~K.,   {Markoff} S.,  2003, \mn@doi [\aap] {10.1051/0004-6361:20030090}, \href {http://cdsads.u-strasbg.fr/abs/2003A%26A...400.1007C} {400, 1007}

\bibitem[\protect\citeauthoryear{{Coti Zelati} et~al.,}{{Coti Zelati} et~al.}{2021}]{cotizelati2021}
{Coti Zelati} F.,  et~al., 2021, \mn@doi [\aap] {10.1051/0004-6361/202141431}, \href {https://ui.adsabs.harvard.edu/abs/2021A&A...655A..52C} {655, A52}

\bibitem[\protect\citeauthoryear{{Court} et~al.,}{{Court} et~al.}{2018}]{court2018}
{Court} J.~M.~C.,  et~al., 2018, \mn@doi [\mnras] {10.1093/mnras/sty2312}, \href {https://ui.adsabs.harvard.edu/abs/2018MNRAS.481.2273C} {481, 2273}

\bibitem[\protect\citeauthoryear{{Courvoisier} et~al.,}{{Courvoisier} et~al.}{2003}]{Courvoisier2003}
{Courvoisier} T.~J.~L.,  et~al., 2003, \mn@doi [\aap] {10.1051/0004-6361:20031172}, \href {https://ui.adsabs.harvard.edu/abs/2003A&A...411L..53C} {411, L53}

\bibitem[\protect\citeauthoryear{{D'Angelo} \& {Spruit}}{{D'Angelo} \& {Spruit}}{2010}]{dangelo2010}
{D'Angelo} C.~R.,  {Spruit} H.~C.,  2010, \mn@doi [\mnras] {10.1111/j.1365-2966.2010.16749.x}, \href {https://ui.adsabs.harvard.edu/abs/2010MNRAS.406.1208D} {406, 1208}

\bibitem[\protect\citeauthoryear{{Das} \& {Porth}}{{Das} \& {Porth}}{2024}]{das2024}
{Das} P.,  {Porth} O.,  2024, \mn@doi [\apjl] {10.3847/2041-8213/ad151f}, \href {https://ui.adsabs.harvard.edu/abs/2024ApJ...960L..12D} {960, L12}

\bibitem[\protect\citeauthoryear{{Das}, {Porth}  \& {Watts}}{{Das} et~al.}{2022}]{das2022}
{Das} P.,  {Porth} O.,   {Watts} A.~L.,  2022, \mn@doi [\mnras] {10.1093/mnras/stac1817}, \href {https://ui.adsabs.harvard.edu/abs/2022MNRAS.515.3144D} {515, 3144}

\bibitem[\protect\citeauthoryear{{Degenaar}, {Miller}, {Harrison}, {Kennea}, {Kouveliotou}  \& {Younes}}{{Degenaar} et~al.}{2014}]{degenaar2014}
{Degenaar} N.,  {Miller} J.~M.,  {Harrison} F.~A.,  {Kennea} J.~A.,  {Kouveliotou} C.,   {Younes} G.,  2014, \mn@doi [\apjl] {10.1088/2041-8205/796/1/L9}, \href {https://ui.adsabs.harvard.edu/abs/2014ApJ...796L...9D} {796, L9}

\bibitem[\protect\citeauthoryear{{Degenaar} et~al.,}{{Degenaar} et~al.}{2018}]{degenaar2018}
{Degenaar} N.,  et~al., 2018, \mn@doi [\ssr] {10.1007/s11214-017-0448-3}, \href {https://ui.adsabs.harvard.edu/abs/2018SSRv..214...15D} {214, 15}

\bibitem[\protect\citeauthoryear{{Di Salvo}, {Papitto}, {Marino}, {Iaria}  \& {Burderi}}{{Di Salvo} et~al.}{2023}]{disalvo2023}
{Di Salvo} T.,  {Papitto} A.,  {Marino} A.,  {Iaria} R.,   {Burderi} L.,  2023, \mn@doi [arXiv e-prints] {10.48550/arXiv.2311.12516}, \href {https://ui.adsabs.harvard.edu/abs/2023arXiv231112516D} {p. arXiv:2311.12516}

\bibitem[\protect\citeauthoryear{{Evans} et~al.,}{{Evans} et~al.}{2007}]{evans2007}
{Evans} P.~A.,  et~al., 2007, \mn@doi [\aap] {10.1051/0004-6361:20077530}, \href {https://ui.adsabs.harvard.edu/abs/2007A&A...469..379E} {469, 379}

\bibitem[\protect\citeauthoryear{{Fender} \& {Hendry}}{{Fender} \& {Hendry}}{2000}]{fender2000}
{Fender} R.~P.,  {Hendry} M.~A.,  2000, \mn@doi [\mnras] {10.1046/j.1365-8711.2000.03443.x}, \href {https://ui.adsabs.harvard.edu/abs/2000MNRAS.317....1F} {317, 1}

\bibitem[\protect\citeauthoryear{{Fender} \& {Kuulkers}}{{Fender} \& {Kuulkers}}{2001}]{fender2001}
{Fender} R.~P.,  {Kuulkers} E.,  2001, \mn@doi [\mnras] {10.1046/j.1365-8711.2001.04345.x}, \href {https://ui.adsabs.harvard.edu/abs/2001MNRAS.324..923F} {324, 923}

\bibitem[\protect\citeauthoryear{{Fender}, {Belloni}  \& {Gallo}}{{Fender} et~al.}{2004}]{fender2004outburststates}
{Fender} R.~P.,  {Belloni} T.~M.,   {Gallo} E.,  2004, \mn@doi [\mnras] {10.1111/j.1365-2966.2004.08384.x}, \href {https://ui.adsabs.harvard.edu/abs/2004MNRAS.355.1105F} {355, 1105}

\bibitem[\protect\citeauthoryear{{Fomalont}, {Geldzahler}  \& {Bradshaw}}{{Fomalont} et~al.}{2001a}]{fomalont2001a}
{Fomalont} E.~B.,  {Geldzahler} B.~J.,   {Bradshaw} C.~F.,  2001a, \mn@doi [\apjl] {10.1086/320490}, \href {https://ui.adsabs.harvard.edu/abs/2001ApJ...553L..27F} {553, L27}

\bibitem[\protect\citeauthoryear{{Fomalont}, {Geldzahler}  \& {Bradshaw}}{{Fomalont} et~al.}{2001b}]{fomalont2001b}
{Fomalont} E.~B.,  {Geldzahler} B.~J.,   {Bradshaw} C.~F.,  2001b, \mn@doi [\apj] {10.1086/322479}, \href {https://ui.adsabs.harvard.edu/abs/2001ApJ...558..283F} {558, 283}

\bibitem[\protect\citeauthoryear{{Fortin}, {Kalsi}, {Garc{\'\i}a}, {Simaz-Bunzel}  \& {Chaty}}{{Fortin} et~al.}{2024}]{fortin2024}
{Fortin} F.,  {Kalsi} A.,  {Garc{\'\i}a} F.,  {Simaz-Bunzel} A.,   {Chaty} S.,  2024, \mn@doi [\aap] {10.1051/0004-6361/202347908}, \href {https://ui.adsabs.harvard.edu/abs/2024A&A...684A.124F} {684, A124}

\bibitem[\protect\citeauthoryear{{Fragile}, {Ballantyne}, {Maccarone}  \& {Witry}}{{Fragile} et~al.}{2018}]{fragile2018}
{Fragile} P.~C.,  {Ballantyne} D.~R.,  {Maccarone} T.~J.,   {Witry} J. W.~L.,  2018, \mn@doi [\apjl] {10.3847/2041-8213/aaeb99}, \href {https://ui.adsabs.harvard.edu/abs/2018ApJ...867L..28F} {867, L28}

\bibitem[\protect\citeauthoryear{{Fragile}, {Ballantyne}  \& {Blankenship}}{{Fragile} et~al.}{2020}]{fragile2020}
{Fragile} P.~C.,  {Ballantyne} D.~R.,   {Blankenship} A.,  2020, \mn@doi [Nature Astronomy] {10.1038/s41550-019-0987-5}, \href {https://ui.adsabs.harvard.edu/abs/2020NatAs...4..541F} {4, 541}

\bibitem[\protect\citeauthoryear{{Frank} et~al.,}{{Frank} et~al.}{2014}]{stellarjetreview2}
{Frank} A.,  et~al., 2014, in {Beuther} H.,  {Klessen} R.~S.,  {Dullemond} C.~P.,   {Henning} T.,  eds, Protostars and Planets VI. pp 451--474 (\mn@eprint {arXiv} {1402.3553}), \mn@doi{10.2458/azu_uapress_9780816531240-ch020}

\bibitem[\protect\citeauthoryear{{Fruchter} \& {Goss}}{{Fruchter} \& {Goss}}{1995}]{fruchter1995}
{Fruchter} A.~S.,  {Goss} W.~M.,  1995, \mn@doi [Journal of Astrophysics and Astronomy] {10.1007/BF02714838}, \href {https://ui.adsabs.harvard.edu/abs/1995JApA...16..245F} {16, 245}

\bibitem[\protect\citeauthoryear{{Gallo}, {Fender}  \& {Pooley}}{{Gallo} et~al.}{2003}]{gallo03}
{Gallo} E.,  {Fender} R.~P.,   {Pooley} G.~G.,  2003, \mn@doi [\mnras] {10.1046/j.1365-8711.2003.06791.x}, \href {http://adsabs.harvard.edu/abs/2003MNRAS.344...60G} {344, 60}

\bibitem[\protect\citeauthoryear{{Gallo}, {Fender}, {Miller-Jones}, {Merloni}, {Jonker}, {Heinz}, {Maccarone}  \& {van der Klis}}{{Gallo} et~al.}{2006}]{gallo2006}
{Gallo} E.,  {Fender} R.~P.,  {Miller-Jones} J.~C.~A.,  {Merloni} A.,  {Jonker} P.~G.,  {Heinz} S.,  {Maccarone} T.~J.,   {van der Klis} M.,  2006, \mn@doi [\mnras] {10.1111/j.1365-2966.2006.10560.x}, \href {https://ui.adsabs.harvard.edu/abs/2006MNRAS.370.1351G} {370, 1351}

\bibitem[\protect\citeauthoryear{{Gallo}, {Degenaar}  \& {van den Eijnden}}{{Gallo} et~al.}{2018}]{gallo18}
{Gallo} E.,  {Degenaar} N.,   {van den Eijnden} J.,  2018, \mn@doi [\mnras] {10.1093/mnrasl/sly083}, \href {http://cdsads.u-strasbg.fr/abs/2018MNRAS.478L.132G} {478, L132}

\bibitem[\protect\citeauthoryear{{Galloway} \& {Keek}}{{Galloway} \& {Keek}}{2021}]{galloway2021}
{Galloway} D.~K.,  {Keek} L.,  2021, in {Belloni} T.~M.,  {M{\'e}ndez} M.,   {Zhang} C.,  eds,  Astrophysics and Space Science Library Vol. 461, Timing Neutron Stars: Pulsations, Oscillations and Explosions. pp 209--262 (\mn@eprint {arXiv} {1712.06227}), \mn@doi{10.1007/978-3-662-62110-3_5}

\bibitem[\protect\citeauthoryear{{Gehrels} et~al.,}{{Gehrels} et~al.}{2004}]{gehrels2004}
{Gehrels} N.,  et~al., 2004, \mn@doi [\apj] {10.1086/422091}, \href {https://ui.adsabs.harvard.edu/abs/2004ApJ...611.1005G} {611, 1005}

\bibitem[\protect\citeauthoryear{{Ghosh} \& {Lamb}}{{Ghosh} \& {Lamb}}{1978}]{ghosh1978}
{Ghosh} P.,  {Lamb} F.~K.,  1978, \mn@doi [\apjl] {10.1086/182734}, \href {https://ui.adsabs.harvard.edu/abs/1978ApJ...223L..83G} {223, L83}

\bibitem[\protect\citeauthoryear{{Grindlay} \& {Seaquist}}{{Grindlay} \& {Seaquist}}{1986}]{grindlay1986}
{Grindlay} J.~E.,  {Seaquist} E.~R.,  1986, \mn@doi [\apj] {10.1086/164673}, \href {https://ui.adsabs.harvard.edu/abs/1986ApJ...310..172G} {310, 172}

\bibitem[\protect\citeauthoryear{{Guerriero} et~al.,}{{Guerriero} et~al.}{1999}]{guerriero1999}
{Guerriero} R.,  et~al., 1999, \mn@doi [\mnras] {10.1046/j.1365-8711.1999.02651.x}, \href {https://ui.adsabs.harvard.edu/abs/1999MNRAS.307..179G} {307, 179}

\bibitem[\protect\citeauthoryear{{Gusinskaia} et~al.,}{{Gusinskaia} et~al.}{2020a}]{gusinskaia2020b}
{Gusinskaia} N.~V.,  et~al., 2020a, \mn@doi [\mnras] {10.1093/mnras/stz3460}, \href {https://ui.adsabs.harvard.edu/abs/2020MNRAS.492.1091G} {492, 1091}

\bibitem[\protect\citeauthoryear{{Gusinskaia} et~al.,}{{Gusinskaia} et~al.}{2020b}]{gusinskaia2020}
{Gusinskaia} N.~V.,  et~al., 2020b, \mn@doi [\mnras] {10.1093/mnras/stz3420}, \href {https://ui.adsabs.harvard.edu/abs/2020MNRAS.492.2858G} {492, 2858}

\bibitem[\protect\citeauthoryear{{Hannikainen}, {Hunstead}, {Campbell-Wilson}  \& {Sood}}{{Hannikainen} et~al.}{1998}]{hannikainen98}
{Hannikainen} D.~C.,  {Hunstead} R.~W.,  {Campbell-Wilson} D.,   {Sood} R.~K.,  1998, \aap, \href {https://ui.adsabs.harvard.edu/abs/1998A&A...337..460H} {337, 460}

\bibitem[\protect\citeauthoryear{{Hjellming}, {Han}, {Cordova}  \& {Hasinger}}{{Hjellming} et~al.}{1990a}]{hjellming1990a}
{Hjellming} R.~M.,  {Han} X.~H.,  {Cordova} F.~A.,   {Hasinger} G.,  1990a, \aap, \href {https://ui.adsabs.harvard.edu/abs/1990A&A...235..147H} {235, 147}

\bibitem[\protect\citeauthoryear{{Hjellming} et~al.,}{{Hjellming} et~al.}{1990b}]{hjellming1990b}
{Hjellming} R.~M.,  et~al., 1990b, \mn@doi [\apj] {10.1086/169522}, \href {https://ui.adsabs.harvard.edu/abs/1990ApJ...365..681H} {365, 681}

\bibitem[\protect\citeauthoryear{{Homer}, {Deutsch}, {Anderson}  \& {Margon}}{{Homer} et~al.}{2001}]{homer2001}
{Homer} L.,  {Deutsch} E.~W.,  {Anderson} S.~F.,   {Margon} B.,  2001, \mn@doi [\aj] {10.1086/323545}, \href {https://ui.adsabs.harvard.edu/abs/2001AJ....122.2627H} {122, 2627}

\bibitem[\protect\citeauthoryear{{Illarionov} \& {Sunyaev}}{{Illarionov} \& {Sunyaev}}{1975}]{illarionov1975}
{Illarionov} A.~F.,  {Sunyaev} R.~A.,  1975, \aap, \href {https://ui.adsabs.harvard.edu/abs/1975A&A....39..185I} {39, 185}

\bibitem[\protect\citeauthoryear{{Johnson}, {Catura}, {Lamb}, {White}, {Sanford}, {Hoffman}, {Lewin}  \& {Jernigan}}{{Johnson} et~al.}{1978}]{johnson1978}
{Johnson} H.~M.,  {Catura} R.~C.,  {Lamb} P.~A.,  {White} N.~E.,  {Sanford} P.~W.,  {Hoffman} J.~A.,  {Lewin} W.~H.~G.,   {Jernigan} J.~G.,  1978, \mn@doi [\apj] {10.1086/156184}, \href {https://ui.adsabs.harvard.edu/abs/1978ApJ...222..664J} {222, 664}

\bibitem[\protect\citeauthoryear{{Johnston}, {Kulkarni}  \& {Goss}}{{Johnston} et~al.}{1991}]{johnston1991}
{Johnston} H.~M.,  {Kulkarni} S.~R.,   {Goss} W.~M.,  1991, \mn@doi [\apjl] {10.1086/186219}, \href {https://ui.adsabs.harvard.edu/abs/1991ApJ...382L..89J} {382, L89}

\bibitem[\protect\citeauthoryear{{Kajava}, {S{\'a}nchez-Fern{\'a}ndez}, {Kuulkers}  \& {Poutanen}}{{Kajava} et~al.}{2017}]{kajava2017}
{Kajava} J.~J.~E.,  {S{\'a}nchez-Fern{\'a}ndez} C.,  {Kuulkers} E.,   {Poutanen} J.,  2017, \mn@doi [\aap] {10.1051/0004-6361/201629542}, \href {https://ui.adsabs.harvard.edu/abs/2017A&A...599A..89K} {599, A89}

\bibitem[\protect\citeauthoryear{{Kumar} \& {Zhang}}{{Kumar} \& {Zhang}}{2015}]{kumar2015}
{Kumar} P.,  {Zhang} B.,  2015, \mn@doi [\physrep] {10.1016/j.physrep.2014.09.008}, \href {https://ui.adsabs.harvard.edu/abs/2015PhR...561....1K} {561, 1}

\bibitem[\protect\citeauthoryear{{Ludlam}}{{Ludlam}}{2024}]{ludlam2024}
{Ludlam} R.~M.,  2024, \mn@doi [\apss] {10.1007/s10509-024-04281-y}, \href {https://ui.adsabs.harvard.edu/abs/2024Ap&SS.369...16L} {369, 16}

\bibitem[\protect\citeauthoryear{{Lund} et~al.,}{{Lund} et~al.}{2003}]{Lund03}
{Lund} N.,  et~al., 2003, \mn@doi [\aap] {10.1051/0004-6361:20031358}, \href {http://adsabs.harvard.edu/abs/2003A%26A...411L.231L} {411, L231}

\bibitem[\protect\citeauthoryear{{Lyutikov}}{{Lyutikov}}{2023}]{lyutikov2023}
{Lyutikov} M.,  2023, \mn@doi [\mnras] {10.1093/mnras/stad284}, \href {https://ui.adsabs.harvard.edu/abs/2023MNRAS.520.4315L} {520, 4315}

\bibitem[\protect\citeauthoryear{{Matsuoka} et~al.,}{{Matsuoka} et~al.}{2009}]{matsuoka2009}
{Matsuoka} M.,  et~al., 2009, \mn@doi [\pasj] {10.1093/pasj/61.5.999}, \href {https://ui.adsabs.harvard.edu/abs/2009PASJ...61..999M} {61, 999}

\bibitem[\protect\citeauthoryear{{Migliari} \& {Fender}}{{Migliari} \& {Fender}}{2006}]{migliari06}
{Migliari} S.,  {Fender} R.~P.,  2006, \mn@doi [\mnras] {10.1111/j.1365-2966.2005.09777.x}, \href {http://adsabs.harvard.edu/abs/2006MNRAS.366...79M} {366, 79}

\bibitem[\protect\citeauthoryear{{Mirabel} \& {Rodr{\'\i}guez}}{{Mirabel} \& {Rodr{\'\i}guez}}{1994}]{mirabel1994}
{Mirabel} I.~F.,  {Rodr{\'\i}guez} L.~F.,  1994, \mn@doi [\nat] {10.1038/371046a0}, \href {https://ui.adsabs.harvard.edu/abs/1994Natur.371...46M} {371, 46}

\bibitem[\protect\citeauthoryear{{Moore}, {Rutledge}, {Fox}, {Guerriero}, {Lewin}, {Fender}  \& {van Paradijs}}{{Moore} et~al.}{2000}]{moore2000}
{Moore} C.~B.,  {Rutledge} R.~E.,  {Fox} D.~W.,  {Guerriero} R.~A.,  {Lewin} W. H.~G.,  {Fender} R.,   {van Paradijs} J.,  2000, \mn@doi [\apj] {10.1086/308589}, \href {https://ui.adsabs.harvard.edu/abs/2000ApJ...532.1181M} {532, 1181}

\bibitem[\protect\citeauthoryear{{Motta} \& {Fender}}{{Motta} \& {Fender}}{2019}]{motta2019}
{Motta} S.~E.,  {Fender} R.~P.,  2019, \mn@doi [\mnras] {10.1093/mnras/sty3331}, \href {https://ui.adsabs.harvard.edu/abs/2019MNRAS.483.3686M} {483, 3686}

\bibitem[\protect\citeauthoryear{{Murguia-Berthier}, {Parfrey}, {Tchekhovskoy}  \& {Jacquemin-Ide}}{{Murguia-Berthier} et~al.}{2024}]{berthier2024}
{Murguia-Berthier} A.,  {Parfrey} K.,  {Tchekhovskoy} A.,   {Jacquemin-Ide} J.,  2024, \mn@doi [\apjl] {10.3847/2041-8213/ad16eb}, \href {https://ui.adsabs.harvard.edu/abs/2024ApJ...961L..20M} {961, L20}

\bibitem[\protect\citeauthoryear{{Papitto} \& {de Martino}}{{Papitto} \& {de Martino}}{2022}]{papitto2022}
{Papitto} A.,  {de Martino} D.,  2022, in {Bhattacharyya} S.,  {Papitto} A.,   {Bhattacharya} D.,  eds,  Astrophysics and Space Science Library Vol. 465, Astrophysics and Space Science Library. pp 157--200 (\mn@eprint {arXiv} {2010.09060}), \mn@doi{10.1007/978-3-030-85198-9_6}

\bibitem[\protect\citeauthoryear{{Parfrey} \& {Tchekhovskoy}}{{Parfrey} \& {Tchekhovskoy}}{2023}]{parfrey2023}
{Parfrey} K.,  {Tchekhovskoy} A.,  2023, \mn@doi [arXiv e-prints] {10.48550/arXiv.2311.04291}, \href {https://ui.adsabs.harvard.edu/abs/2023arXiv231104291P} {p. arXiv:2311.04291}

\bibitem[\protect\citeauthoryear{{Parfrey}, {Spitkovsky}  \& {Beloborodov}}{{Parfrey} et~al.}{2016}]{parfrey2016}
{Parfrey} K.,  {Spitkovsky} A.,   {Beloborodov} A.~M.,  2016, \mn@doi [\apj] {10.3847/0004-637X/822/1/33}, \href {http://adsabs.harvard.edu/abs/2016ApJ...822...33P} {822, 33}

\bibitem[\protect\citeauthoryear{{Penninx}, {Lewin}, {Zijlstra}, {Mitsuda}  \& {van Paradijs}}{{Penninx} et~al.}{1988}]{penninx1988}
{Penninx} W.,  {Lewin} W. H.~G.,  {Zijlstra} A.~A.,  {Mitsuda} K.,   {van Paradijs} J.,  1988, \mn@doi [\nat] {10.1038/336146a0}, \href {https://ui.adsabs.harvard.edu/abs/1988Natur.336..146P} {336, 146}

\bibitem[\protect\citeauthoryear{{Pooley} \& {Fender}}{{Pooley} \& {Fender}}{1997}]{pooley1997}
{Pooley} G.~G.,  {Fender} R.~P.,  1997, \mn@doi [\mnras] {10.1093/mnras/292.4.925}, \href {https://ui.adsabs.harvard.edu/abs/1997MNRAS.292..925P} {292, 925}

\bibitem[\protect\citeauthoryear{{Pudritz}, {Hardcastle}  \& {Gabuzda}}{{Pudritz} et~al.}{2012}]{stellarjetreview1}
{Pudritz} R.~E.,  {Hardcastle} M.~J.,   {Gabuzda} D.~C.,  2012, \mn@doi [\ssr] {10.1007/s11214-012-9895-z}, \href {https://ui.adsabs.harvard.edu/abs/2012SSRv..169...27P} {169, 27}

\bibitem[\protect\citeauthoryear{{Rao} \& {Venugopal}}{{Rao} \& {Venugopal}}{1980}]{rao1980}
{Rao} A.~P.,  {Venugopal} V.~R.,  1980, Bulletin of the Astronomical Society of India, \href {https://ui.adsabs.harvard.edu/abs/1980BASI....8...41R} {8, 41}

\bibitem[\protect\citeauthoryear{{Russell}, {Degenaar}, {Miller-Jones}  \& {Tudor}}{{Russell} et~al.}{2017}]{russell2017}
{Russell} T.,  {Degenaar} N.,  {Miller-Jones} J.,   {Tudor} V.,  2017, The Astronomer's Telegram, \href {https://ui.adsabs.harvard.edu/abs/2017ATel10106....1R} {10106, 1}

\bibitem[\protect\citeauthoryear{{Russell}, {Degenaar}, {Wijnands}, {van den Eijnden}, {Gusinskaia}, {Hessels}  \& {Miller-Jones}}{{Russell} et~al.}{2018}]{russell2018}
{Russell} T.~D.,  {Degenaar} N.,  {Wijnands} R.,  {van den Eijnden} J.,  {Gusinskaia} N.~V.,  {Hessels} J.~W.~T.,   {Miller-Jones} J.~C.~A.,  2018, \mn@doi [\apjl] {10.3847/2041-8213/aaf4f9}, \href {https://ui.adsabs.harvard.edu/abs/2018ApJ...869L..16R} {869, L16}

\bibitem[\protect\citeauthoryear{{Russell} et~al.,}{{Russell} et~al.}{2019}]{russel2019}
{Russell} T.~D.,  et~al., 2019, \mn@doi [\apj] {10.3847/1538-4357/ab3d36}, \href {https://ui.adsabs.harvard.edu/abs/2019ApJ...883..198R} {883, 198}

\bibitem[\protect\citeauthoryear{{Russell} et~al.,}{{Russell} et~al.}{2020}]{russell2020}
{Russell} T.~D.,  et~al., 2020, \mn@doi [\mnras] {10.1093/mnras/staa2650}, \href {https://ui.adsabs.harvard.edu/abs/2020MNRAS.498.5772R} {498, 5772}

\bibitem[\protect\citeauthoryear{{Russell} et~al.,}{{Russell} et~al.}{2024}]{russell2024}
{Russell} T.~D.,  et~al., 2024, \mn@doi [\nat] {10.1038/s41586-024-07133-5}, \href {https://ui.adsabs.harvard.edu/abs/2024Natur.627..763R} {627, 763}

\bibitem[\protect\citeauthoryear{{S{\'a}nchez-Fern{\'a}ndez}, {Kajava}, {Poutanen}, {Kuulkers}  \& {Suleimanov}}{{S{\'a}nchez-Fern{\'a}ndez} et~al.}{2020}]{sanchez2020}
{S{\'a}nchez-Fern{\'a}ndez} C.,  {Kajava} J.~J.~E.,  {Poutanen} J.,  {Kuulkers} E.,   {Suleimanov} V.~F.,  2020, \mn@doi [\aap] {10.1051/0004-6361/201936599}, \href {https://ui.adsabs.harvard.edu/abs/2020A&A...634A..58S} {634, A58}

\bibitem[\protect\citeauthoryear{{Selina} et~al.,}{{Selina} et~al.}{2018}]{selina18}
{Selina} R.~J.,  et~al., 2018, {The ngVLA Reference Design}.
p.~15

\bibitem[\protect\citeauthoryear{{Spruit} \& {Taam}}{{Spruit} \& {Taam}}{1993}]{spruit1993}
{Spruit} H.~C.,  {Taam} R.~E.,  1993, \mn@doi [\apj] {10.1086/172162}, \href {https://ui.adsabs.harvard.edu/abs/1993ApJ...402..593S} {402, 593}

\bibitem[\protect\citeauthoryear{{Tetarenko} et~al.,}{{Tetarenko} et~al.}{2017}]{atetarenko2017}
{Tetarenko} A.~J.,  et~al., 2017, \mn@doi [\mnras] {10.1093/mnras/stx1048}, \href {https://ui.adsabs.harvard.edu/abs/2017MNRAS.469.3141T} {469, 3141}

\bibitem[\protect\citeauthoryear{{Tetarenko}, {Casella}, {Miller-Jones}, {Sivakoff}, {Tetarenko}, {Maccarone}, {Gandhi}  \& {Eikenberry}}{{Tetarenko} et~al.}{2019}]{atetarenko2019_cygx1}
{Tetarenko} A.~J.,  {Casella} P.,  {Miller-Jones} J.~C.~A.,  {Sivakoff} G.~R.,  {Tetarenko} B.~E.,  {Maccarone} T.~J.,  {Gandhi} P.,   {Eikenberry} S.,  2019, \mn@doi [\mnras] {10.1093/mnras/stz165}, \href {https://ui.adsabs.harvard.edu/abs/2019MNRAS.484.2987T} {484, 2987}

\bibitem[\protect\citeauthoryear{{Tetarenko} et~al.,}{{Tetarenko} et~al.}{2021}]{atetarenko2021}
{Tetarenko} A.~J.,  et~al., 2021, \mn@doi [\mnras] {10.1093/mnras/stab820}, \href {https://ui.adsabs.harvard.edu/abs/2021MNRAS.504.3862T} {504, 3862}

\bibitem[\protect\citeauthoryear{{Tudor} et~al.,}{{Tudor} et~al.}{2017}]{tudor17}
{Tudor} V.,  et~al., 2017, \mn@doi [\mnras] {10.1093/mnras/stx1168}, \href {http://adsabs.harvard.edu/abs/2017MNRAS.470..324T} {470, 324}

\bibitem[\protect\citeauthoryear{{Tudor} et~al.,}{{Tudor} et~al.}{2022}]{tudor2022}
{Tudor} V.,  et~al., 2022, \mn@doi [\mnras] {10.1093/mnras/stac1034}, \href {https://ui.adsabs.harvard.edu/abs/2022MNRAS.513.3818T} {513, 3818}

\bibitem[\protect\citeauthoryear{{Vadawale}, {Rao}, {Naik}, {Yadav}, {Ishwara-Chandra}, {Pramesh Rao}  \& {Pooley}}{{Vadawale} et~al.}{2003}]{vadawale2003}
{Vadawale} S.~V.,  {Rao} A.~R.,  {Naik} S.,  {Yadav} J.~S.,  {Ishwara-Chandra} C.~H.,  {Pramesh Rao} A.,   {Pooley} G.~G.,  2003, \mn@doi [\apj] {10.1086/378672}, \href {https://ui.adsabs.harvard.edu/abs/2003ApJ...597.1023V} {597, 1023}

\bibitem[\protect\citeauthoryear{{Valenti}, {Ferraro}  \& {Origlia}}{{Valenti} et~al.}{2010}]{valenti2010}
{Valenti} E.,  {Ferraro} F.~R.,   {Origlia} L.,  2010, \mn@doi [\mnras] {10.1111/j.1365-2966.2009.15991.x}, \href {https://ui.adsabs.harvard.edu/abs/2010MNRAS.402.1729V} {402, 1729}

\bibitem[\protect\citeauthoryear{{Vaughan}, {Edelson}, {Warwick}  \& {Uttley}}{{Vaughan} et~al.}{2003}]{vaughan2003}
{Vaughan} S.,  {Edelson} R.,  {Warwick} R.~S.,   {Uttley} P.,  2003, \mn@doi [\mnras] {10.1046/j.1365-2966.2003.07042.x}, \href {https://ui.adsabs.harvard.edu/abs/2003MNRAS.345.1271V} {345, 1271}

\bibitem[\protect\citeauthoryear{{Verner}, {Ferland}, {Korista}  \& {Yakovlev}}{{Verner} et~al.}{1996}]{vern1996}
{Verner} D.~A.,  {Ferland} G.~J.,  {Korista} K.~T.,   {Yakovlev} D.~G.,  1996, \mn@doi [\apj] {10.1086/177435}, \href {https://ui.adsabs.harvard.edu/abs/1996ApJ...465..487V} {465, 487}

\bibitem[\protect\citeauthoryear{{Vincentelli} et~al.,}{{Vincentelli} et~al.}{2023}]{vincentelli2023}
{Vincentelli} F.~M.,  et~al., 2023, \mn@doi [\nat] {10.1038/s41586-022-05648-3}, \href {https://ui.adsabs.harvard.edu/abs/2023Natur.615...45V} {615, 45}

\bibitem[\protect\citeauthoryear{{Wilms}, {Allen}  \& {McCray}}{{Wilms} et~al.}{2000}]{wilms2000}
{Wilms} J.,  {Allen} A.,   {McCray} R.,  2000, \mn@doi [\apj] {10.1086/317016}, \href {https://ui.adsabs.harvard.edu/abs/2000ApJ...542..914W} {542, 914}

\bibitem[\protect\citeauthoryear{{Winkler} et~al.,}{{Winkler} et~al.}{2003}]{winkler2003}
{Winkler} C.,  et~al., 2003, \mn@doi [\aap] {10.1051/0004-6361:20031288}, \href {https://ui.adsabs.harvard.edu/abs/2003A&A...411L...1W} {411, L1}

\bibitem[\protect\citeauthoryear{{Wood} et~al.,}{{Wood} et~al.}{2021}]{wood2021}
{Wood} C.~M.,  et~al., 2021, \mn@doi [\mnras] {10.1093/mnras/stab1479}, \href {https://ui.adsabs.harvard.edu/abs/2021MNRAS.505.3393W} {505, 3393}

\bibitem[\protect\citeauthoryear{{Younes} et~al.,}{{Younes} et~al.}{2015}]{younes2015}
{Younes} G.,  et~al., 2015, \mn@doi [\apj] {10.1088/0004-637X/804/1/43}, \href {https://ui.adsabs.harvard.edu/abs/2015ApJ...804...43Y} {804, 43}

\bibitem[\protect\citeauthoryear{{van den Eijnden}, {Bagnoli}, {Degenaar}, {Lohfink}, {Parker}, {in 't Zand}  \& {Fabian}}{{van den Eijnden} et~al.}{2017}]{vandeneijnden2017}
{van den Eijnden} J.,  {Bagnoli} T.,  {Degenaar} N.,  {Lohfink} A.~M.,  {Parker} M.~L.,  {in 't Zand} J.~J.~M.,   {Fabian} A.~C.,  2017, \mn@doi [\mnras] {10.1093/mnrasl/slw244}, \href {https://ui.adsabs.harvard.edu/abs/2017MNRAS.466L..98V} {466, L98}

\bibitem[\protect\citeauthoryear{{van den Eijnden} et~al.,}{{van den Eijnden} et~al.}{2020}]{vandeneijnden2020}
{van den Eijnden} J.,  et~al., 2020, \mn@doi [\mnras] {10.1093/mnras/staa1704}, \href {https://ui.adsabs.harvard.edu/abs/2020MNRAS.496.4127V} {496, 4127}

\bibitem[\protect\citeauthoryear{{van den Eijnden} et~al.,}{{van den Eijnden} et~al.}{2021}]{vandeneijnden2021}
{van den Eijnden} J.,  et~al., 2021, \mn@doi [\mnras] {10.1093/mnras/stab1995}, \href {https://ui.adsabs.harvard.edu/abs/2021MNRAS.507.3899V} {507, 3899}

\makeatother
\end{thebibliography}

%%%%%%%%%%%%%%%%%%%%%%%%%%%%%%%%%%%%%%%%%%%%%%%%%%

% Don't change these lines
\bsp	% typesetting comment
\label{lastpage}
\end{document}